\documentclass[aps,pra,reprint,10pt,a4paper]{revtex4-2}
\usepackage[utf8]{inputenc}
\usepackage[sc,osf]{mathpazo}
\usepackage{amsmath}
\usepackage[T1]{fontenc}
\usepackage{latexsym}
\usepackage{amssymb}
\usepackage[colorlinks=true,citecolor=blue,urlcolor=blue]{hyperref}
\usepackage{color}
\usepackage{graphics,epstopdf}
\usepackage{soul}
\usepackage[demo]{graphicx}
\usepackage{capt-of}
\usepackage{lipsum}
\usepackage{adjustbox}
\usepackage[normalem]{ulem}
\usepackage[table,xcdraw]{xcolor}
\usepackage{braket}
\usepackage{physics}
\usepackage{ragged2e}

\usepackage{comment}
\usepackage{amsmath}
\usepackage{graphicx}
\usepackage{orcidlink}
\usepackage[font=small,labelfont=bf,justification=justified,singlelinecheck=false]{caption}

\newcommand{\ms}[1]{{\color{black}{{#1}}}}

\begin{document}

\title{Continuous variable dense coding under realistic non-ideal scenarios\\
}

\author{Mrinmoy Samanta$^{1, 2}$\,\orcidlink{0009-0009-8406-731X}, Ayan Patra$^{1, 2}$\,\orcidlink{0000-0002-7761-5623}, Rivu Gupta$^{1, 2, 3}$\,\orcidlink{0000-0002-0254-2028}, Aditi Sen(De)$^{1, 2}$\,\orcidlink{0000-0003-1693-0440}}

\affiliation{$^1$  Harish-Chandra Research Institute, Chhatnag Road, Jhunsi, Allahabad - \(211019\), India}

\affiliation{$^2$ Homi Bhabha National Institute, Training School Complex, Anushakti Nagar, Mumbai \(400 094\), India}

\affiliation{$^3$ Dipartimento di Fisica “Aldo Pontremoli,” Università degli Studi di Milano, I-$20133$ Milano, Italy}

\begin{abstract}


We analyze the continuous variable (CV)  dense coding protocol between a single sender and a single receiver when affected by noise in the shared and encoded states, as well as when the decoding is imperfect. \ms{We present a formalism to derive the dense coding capacity (DCC) of generic two-mode Gaussian states when affected by Gaussian noise models.} 
When the constituent modes are affected by quantum-limited amplifiers, pure-loss channels, and environmental interactions, together with an inefficient decoding mechanism comprising imperfect double-homodyne detection, we specifically investigate the performance of the two-mode squeezed vacuum state (TMSV) by varying the strength of the noise. Through numerical simulations, we further show that \ms{the negative conditional entropy may be used to assess the quantum advantage of CV dense coding in Gaussian setups.} While comparing the TMSV with other Gaussian states, we identify a class of pure \ms{Gaussian states capable of furnishing a capacity equal to that of the TMSV, under equal energy constraints in the ideal regime.} Further, we demonstrate that, while the TMSV state is \ms{more susceptible to noise}, it still provides the maximum quantum advantage in the DC protocol under practical conditions.

\end{abstract}

\maketitle

\section{Introduction}
\label{sec:intro}

Quantum teleportation~\cite{Bennett_1993} and superdense coding~\cite{Bennett_PRL_1992} are among the intriguing quantum communication protocols that have captivated the scientific community, highlighting the remarkable potential of entanglement~\cite{Horodecki_RMP_2009}. While the first allows for transmitting quantum information from one place to another distant location, the latter involves transmitting classical information from a sender to a receiver
with the aid of pre-shared entanglement~\cite{Bennett_PRL_1992, sen_PN_2010, GisinComm, Dobrzanski,sen_PRA_2003}, which enhances the transmission capacity compared to the optimal classical scheme (for networks, see Refs. \cite{Bruss_PRL_2004, Bruss_IJQI_2006, Das_PRA_2015, Das_PRA_2014, Prabhu}). 
Quantum advantage in dense coding (DC) has been demonstrated using several physical platforms such as trapped ions~\cite{Libfried_RMP_2003, Yang_JPB_2007}, photons~\cite{Mattle_PRL_1996, Shimizu_PRA_1999, Mizuno_PRA_2005, Pan_RMP_2012, Northup_NP_2014, Barreiro_NP_2014, Krenn_PNAS_2016}, and nuclear magnetic resonance~\cite{Fang_PRA_2000, Wei_CSB_2004, Vandersypen_RMP_2005}. 

In practical situations, noise is unavoidable, and systems inevitably get disturbed due to the interaction with the environment, which, in general, degrades quantum properties. For instance, in the presence of local dephasing noise, entanglement can abruptly vanish, a phenomenon known as entanglement sudden death~\cite{Yu_Science_2009}, which, in turn, negatively impacts various information processing tasks, including DC. Deviations from the ideal DC protocol can occur in multiple ways. Noise can interfere during the distribution of resources between the sender(s) and receiver(s)~\cite{Bose, Bowen_PRA_2001, Horodecki_arxiv_2001, Ziman, Bruss_PRL_2004, Bruss_IJQI_2006}. Alternatively, after the encoding process, the encoded quantum state may be transmitted through a noisy channel~\cite{Shadman_NJP_2010, Shadman2011, Shadman_PRA_2012, Shadman2013}. Additionally, the measurement operations performed at the receiver's end for decoding can be imperfect. Numerous studies have been devoted to addressing these situations, investigating the effects of both Markovian~\cite{Das_PRA_2014, Das_PRA_2015, Mirmasoudi_JPA_2018} and non-Markovian noise~\cite{Liu_EPL_2016, Muhuri_PRA_2024} on the {\color{black}DC protocol primarily in finite-dimensional settings. }

Continuous variable (CV) systems, on the other hand, offer an important platform for implementing quantum schemes since they can overcome specific challenges encountered in finite-dimensional systems, including the difficulty in distinguishing Bell states using linear optics~\cite{Lutkenhaus_PRA_1999} during the classical information decoding stage of DC. The exploration of the dense coding protocol within the realm of CV systems was initiated by Ban \cite{ban-1999} and independently by Braunstein and Kimble \cite{Braunstein_PRA_2000}. Their pioneering work entails the sharing of an Einstein-Podolsky-Rosen (EPR)~\cite{Einstein_PRA_1935} state between a single sender and a single receiver for the transmission of classical information in which the sender encodes information using a displacement operator with parameters chosen from a Gaussian distribution with zero mean and a fixed variance, and the receiver decodes the information by applying a double-homodyne measurement (see Ref. \cite{Patra_PRA_2022} for DC with multiple modes).
Following the theoretical proposal, numerous experimental efforts have been undertaken to realize classical information transmission in CV systems, specifically using two-mode Gaussian states \cite{Kim_PRA_2002, Ralph_PRA_2002, Li_PRL_2002, Jing_PRL_2003, Barzanjeh_PRA_2013, Wang_OptEx_2020, Hao_PRL_2021}. However, the study that deals with the effect of noise on the performance of CV dense coding is limited in the literature. Only exceptions include the investigation of the DC scheme in the presence of entanglement impurity and detector inefficiency~\cite{Ralph_PRA_2002}, and when the quantum channel is subjected to linear dissipation~\cite{Ban_JOB_2000}. 

\begin{figure*}[ht]
\includegraphics[width=\linewidth]{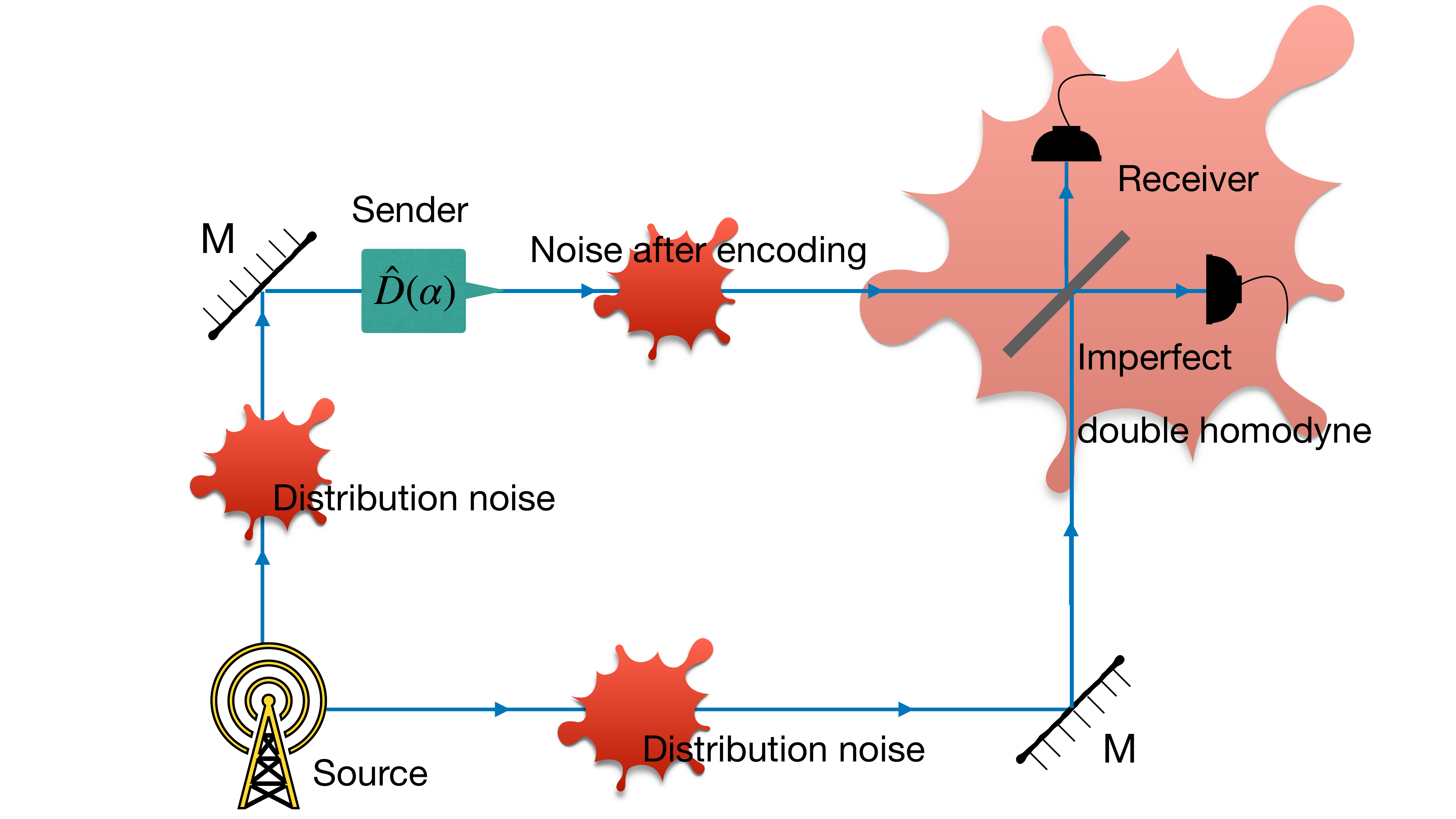}
\captionsetup{justification=Justified,singlelinecheck=false}
\caption{\textbf{Schematic representation of the noisy continuous variable dense coding protocol.} A source emits a two-mode state, one mode being sent to the sender and the other to the receiver. During distribution, noise acts on the two modes, thereby affecting the properties of the state. The sender encodes information via a displacement operation, $\hat{D}(\alpha)$, and transmits the encoded mode to the receiver station. Note that noise also acts on the encoded mode, which travels from the sender to the receiver. The receiver then performs a double-homodyne measurement to decode the message, a process that is also considered to be imperfect. M represents mirrors.}
\label{fig:schematic} 
\end{figure*}

In this work, we examine how the performance of the DC protocol in CV systems gets disturbed in the presence of noise affecting the shared and encoded state as well as the decoding process (see Fig. \ref{fig:schematic}).
We explore the impact of two noisy situations on DC -- $(1)$ \ms{when the parties have knowledge of the specific type of noise affecting the system and tailor the protocol accordingly, which we call the adaptive noisy scenario, and $(2)$ the non-adaptive case, where the parties are unaware of the specific noise model and proceed assuming ideal conditions.} In both pictures, we report the expressions for the DC capacity under the energy constraint at the sender's end when the shared state is a generic two-mode Gaussian state. Our findings reveal that the adaptive scheme consistently yields a higher dense coding capacity and extends the range of noise strengths over which quantum advantage is maintained in comparison with the non-adaptive one. \textcolor{black}{Note that, for discrete variable systems, the adaptive scheme has been employed to derive the channel capacity of Pauli channels in arbitrary finite dimensions~\cite{Laurenza_PRR_2020_channel-capacity} when the noisy channel is present both during resource distribution as well as after encoding. However, in such discrete-variable scenarios, the Holevo quantity~\cite{Holevo-1998} is maximized over the encoding scheme, whereas, due to the energy constraints that must be imposed in CV setups, we must optimize the mutual information between the sender and the receiver and cannot apply the same technique to the CV version of the Holevo quantity~\cite{Braunstein_PRA_2000}. In our work, we \ms{provide a formalism for estimating the dense coding capacity of two-mode Gaussian states}, instead of focusing on the capacity of the noisy channels, and analyze the effects of various realistic imperfections.}

For demonstration, we examine two exemplary noisy channels, the amplifier and the attenuator channels~\cite{Serafini_2017}. Using the two-mode squeezed vacuum (TMSV) state as a resource, we determine the threshold noise strengths for both types of noise, below which the system exhibits non-classical capacity.
{\color{black}Interestingly, we observe that even though the TMSV state loses its performance faster when noise is added, it still gives the best quantum advantage compared to other states, showing that it remains the most effective choice even in a noisy environment.
}

In discrete variable systems, the negative conditional entropy of the shared state is recognized as a key resource for achieving quantum advantage in DC. However, in CV systems, this connection is less obvious. Interestingly, \textcolor{black}{our numerical studies suggest that the negative conditional entropy can also be used to track the quantum advantage in Gaussian CV dense coding protocols, both in the noiseless and the noisy scenarios. \color{black}{In our analysis, we consider as exemplary resources, the TMSV state and states obtained after an active local symplectic operation on a generic two-mode Gaussian state and compare their dense coding capacity and the conditional entropy, under various noisy environments.}}

We finally present an example of a class of pure states that can achieve the same dense coding capacity as the TMSV state with equivalent energy, and exhibit that they also possess the same entanglement content. Further, similar to the discrete variable case, we prove a one-to-one correspondence between the Holevo quantity and the entanglement of pure two-mode Gaussian states having the same energy at the sender's mode.

The paper is organized in the following way. \ms{In Sec. \ref{sec:sec2}, we illustrate how the conditional entropy is a precursor to the quantum advantage in Gaussian CV dense coding when conditions are ideal.} Sec.~\ref{sec:noisy_DC_formalism} introduces the noisy dense coding framework, illustrating how noise can be incorporated during both state distribution and the transmission of the encoded mode through a noisy channel. We also address the treatment of imperfect double-homodyne measurement and derive the computation of DC capacity for both adaptive and non-adaptive strategies. \textcolor{black}{ In Sec. \ref{sec:dcc_analysis_exemplarychannel}, we analyze {the dense coding capacity of generic Gaussian states}, with a special emphasis on the TMSV state, by considering some paradigmatic noise models, and \ms{highlight the role of the conditional entropy in predicting quantum advantage even in noisy scenarios.}} Following this, in Sec. \ref{section:tmsv_other_Gaussian}, we compare the performance of the TMSV state against other pure Gaussian states in both noiseless and noisy regimes. While there exists a class of pure states capable of achieving the same dense coding capacity as the TMSV state with equivalent energy when the protocol operates ideally, we demonstrate a peculiar behavior of the {TMSV state under decoherence -- it provides the maximum quantum advantage even while being maximally susceptible to noise.} Finally, the concluding remarks are provided in Sec. \ref{sec:conclu}.

\section{Role of conditional entropy in the ideal Gaussian CV dense coding protocol}
\label{sec:sec2}

 Before presenting the noisy DC setup, 
 we briefly describe the noiseless CV dense coding routine, where a two-mode entangled state, such as the TMSV  state~\cite{Einstein_PRA_1935}, is employed as a resource~\cite{Braunstein_PRA_2000}. After the distribution of the TMSV state between the sender, $A$, and the receiver, $B$ by the source, $A$ encodes the classical message, $\alpha$, in his(her) mode through a local displacement operation, with respect to a fixed probability distribution of messages, $P(\alpha)$, and transmits the encoded mode to $B$ through a noiseless quantum channel. Subsequently, $B$ performs a perfect double-homodyne measurement on both modes at his(her) disposal to recover the message~\cite{Braunstein_PRA_2000}.

 In discrete-variable systems, the performance of DC is measured via the dense coding capacity (DCC), denoted as $\mathbf{C}$, which is obtained by optimizing the Holevo quantity~\cite{Holevo1973, Holevo-1998, Hiroshima_JPA_2001} (see Appendix \ref{app:numerics} for the role of the Holevo quantity in CV systems). On the other hand, in the CV formalism, the DCC depends on the energy at the sender's side of the shared resource, equivalent to the number of photons in the sender's mode. For a fixed photon number at the sender's side, $\bar{N}_S$, the noiseless dense coding capacity of the TMSV state is given by $\log(1+ \bar{N}_S + \bar{N}_S^2)$~\cite{Braunstein_PRA_2000} (unless otherwise specified, throughout the manuscript, $\log$ refers to the natural logarithm). To determine the success of any information-theoretic protocol, one must compare its figure of merit with that of the corresponding classical scheme. \textcolor{black}{In this case, the best classical protocol comprises a collection of coherent states, whose amplitudes are chosen from a Gaussian distribution, thereby yielding a capacity of $\mathbf{C}_{\text{cl}}=(\bar{N}_S+1)\log (\bar{N}_S+1)-\bar{N}_S\log \bar{N}_S$~\cite{Yuen_Ozawa-1993, drummond_Caves-1994}. We claim that the quantum advantage, $\mathcal{Q}$, is achieved when the capacity of the DC protocol surpasses the classical threshold, i.e., when $\mathcal{Q} = \mathbf{C} - \mathbf{C}_{\text{cl}} > 0$.}
\begin{figure*}[ht]
\includegraphics[width=1\linewidth]{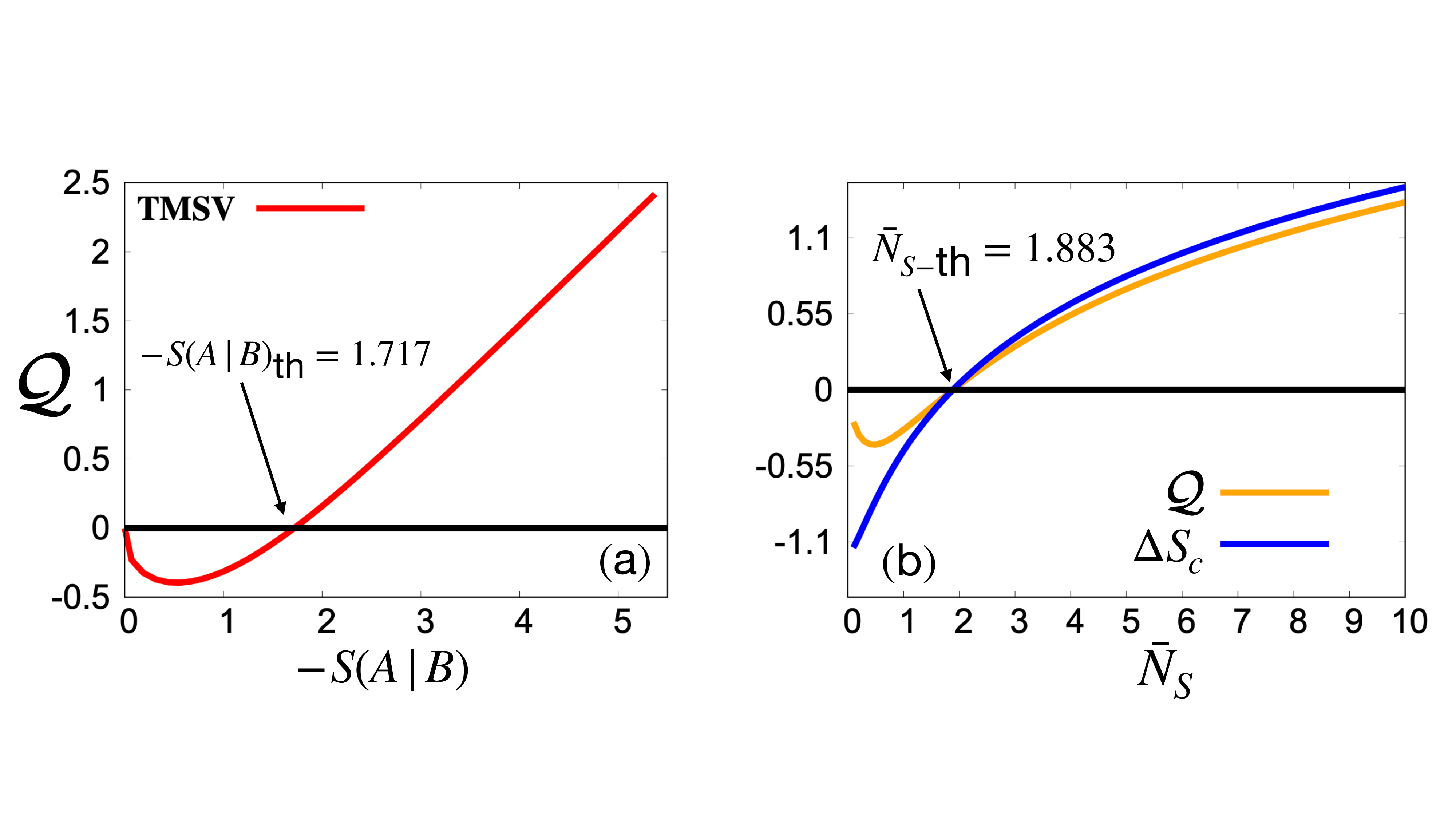}
\captionsetup{justification=Justified,singlelinecheck=false}
\caption{\textbf{Quantum advantage and negative conditional entropy for the noiseless TMSV state.} (a). The variation of the quantum advantage $\mathcal{Q}$ (ordinate) vs the negative conditional entropy $-S(A|B)$ (abscissa) for the noiseless TMSV state. Both axes are represented in bits. (b) The quantum advantage, $\mathcal{Q}$ (light orange line) and $\Delta S_c$ (dark blue line) (ordinate)  with respect to the sender's-side energy, $\bar{N}_S$ (abscissa). The y-axis is in bits while the x-axis is dimensionless. }
\label{fig:noiseless-TMSV-cond-ent} 
\end{figure*}
\textcolor{black}{Having established the criterion for quantum advantage, the fundamental question is, {\it what physical resource enables this excess capacity beyond the classical limit?}} In discrete variable settings, the negative conditional entropy is known to provide quantum advantage in the dense coding protocol ~\cite{Bruss_IJQI_2006}, while in CV systems, the resource responsible for quantum advantage is still unclear. \ms{We shall argue here that the negative conditional entropy can be used to explain the quantum advantage of the dense coding protocol in the CV paradigm using Gaussian states.}

The conditional entropy of the sender's mode is defined as $S(A|B) = S(\rho_{AB}) - S(\rho_B)$ with $\rho_B$ being the subsystem of the receiver's mode. For a single sender and a single receiver involved in the protocol, this also represents the negative of the coherent information~\cite{Schumacher_PRA_1996_data-processing, Horodecki_JPA_2012_dense-coding} between the two parties. Given any Gaussian state, $\chi$, the von Neumann entropy is given by $S(\chi) = \sum_{j = 1}^n s (\nu_j)$~\cite{Serafini_2017}, where, $\nu_j$ are the symplectic eigenvalues of the covariance matrix of the state, and 
\begin{equation}
    s(x) = \frac{x+1}{2} \log_2 \frac{x+1}{2} - \frac{x-1}{2} \log_2 \frac{x-1}{2}.
    \label{eq:entropy_fn}
\end{equation}
 For an $N$-mode covariance matrix, $\Xi_N$, the symplectic eigenvalues are given by the positive eigenvalues of $\iota \Omega \Xi_N$ with $\Omega = \oplus_{i = 1}^N \begin{pmatrix}
    0 & 1 \\
    -1 & 0
\end{pmatrix}$ being the symplectic form and $\iota = \sqrt{-1}$.


As mentioned before, noiseless dense coding with the TMSV state has DCC given as, $\mathbf{C}_{\text{TMSV}} = \log (1 + \bar{N}_S + \bar{N}_S^2)$. At the same time, the quantum advantage reads $\mathcal{Q}_{\text{TMSV}}(\bar{N}_S) = \mathbf{C}_{\text{TMSV}} - \mathbf{C}_{\text{cl}}$ which is positive only when the sender-side energy is above a certain threshold value, i.e., $\bar{N}_S > \bar{N}_{S-\text{th}} =  1.883$. On the other hand, the negative conditional entropy has the form
\begin{widetext}
\begin{equation}
    - S(A|B)_{\text{TMSV}}(\bar{N}_S) = \frac{(1 + \bar{N}_S)^2 \log_2 (1 + \bar{N}_S)^2 - \bar{N}_S^2 \log_2 \bar{N}_S^2 - (1 + 2 \bar{N}_S) \log_2 (1 + 2 \bar{N}_S)}{1 + 2 \bar{N}_S}.
    \label{eq:cond_ent_TMSV}
\end{equation}
\end{widetext}

The variation of the quantum advantage, $\mathcal{Q}$, with respect to $-S(A|B)$ for $\bar{N}_S \in (0,30)$ shows that the quantum advantage is positive beyond $-S(A|B)_{\text{th}} = 1.717$ (see Fig. \ref{fig:noiseless-TMSV-cond-ent}(a)), i.e., the resource state must have a definite value of negative conditional entropy in order to overcome the classical threshold value. Interestingly, this threshold conditional entropy provides a limit on the sender-side energy, i.e., -$S(A|B)_{\text{th}} = - S(A|B)_{\text{TMSV}}(\bar{N}_{S-\text{th}})$. \ms{To analyze the role of negative conditional entropy in gauging the quantum advantage, we define the quantity, $\Delta S_c = -S(A|B) + S(A|B)_{\text{th}}$.} \textcolor{black}{In the scenario of discrete variable noiseless dense coding, a resource state offers a quantum advantage if and only if it possesses negative conditional entropy. Therefore, all pure entangled states provide quantum advantage in this context. However, in CV systems, a pure state must have a certain threshold energy to provide a quantum advantage, even if it is entangled. This indicates that achieving quantum advantage requires the resource state to meet this energy threshold, which justifies the subtraction of threshold negative conditional entropy when investigating the resource in CV dense coding.}  \textcolor{black}{It is clear from Fig. \ref{fig:noiseless-TMSV-cond-ent}(b), that beyond the energy threshold, $\bar{N}_{S-\text{th}}$, the quantity $\Delta S_c$ is concomitant with the quantum advantage, and as such, can be used as an indicator of the same in the noiseless CV dense coding scheme using the TMSV state.}
In the following sections, we shall demonstrate that $\Delta S_c$ is synchronal with the adaptive quantum advantage even in the noisy scenario with generic Gaussian states, thereby \textcolor{black}{providing numerical evidence for the negative conditional entropy to be an indicator of quantum advantage in Gaussian CV dense coding schemes.}

\section{Noisy CV dense coding}
\label{sec:noisy_DC_formalism}
 In real-world applications of dense coding, one can seldom avoid noise and imperfections. In this work, we shall consider three different kinds of shortcomings that may affect the dense coding routine - $(1)$ noise disturbing both the modes during the distribution of the resource between the sender and the receiver, $(2)$ noisy channel transmitting the encoded mode, and $(3)$ an imperfect double-homodyne setup during the decoding step, as represented schematically in Fig. \ref{fig:schematic}. Although in finite dimensional systems, noisy dense coding has been widely studied~\cite{Bose, Bowen_PRA_2001, Horodecki_arxiv_2001, Ziman, Bruss_PRL_2004, Bruss_IJQI_2006}, when noise acts both before and after the encoding process~\cite{Quek_PRA_2010, Shadman_NJP_2010, Shadman_PRA_2012, Shadman2013, Das_PRA_2014, Das_PRA_2015, Mirmasoudi_JPA_2018}, \textcolor{black}{the effect of noise in CV dense coding has not been addressed in full generality, which is the main aim of this paper.}

\textcolor{black}{To analyze the effect of these imperfections in a transparent and unified manner, we adopt the phase-space description, in which a Gaussian state is fully specified by its displacement vector \(\mathbf d\) and covariance matrix \(\Xi\). Up to local symplectic operations, a generic two-mode Gaussian state, $\rho_{AB},$ is characterized as~\cite{Duan_PRL_2000, Simon_PRL_2000, Adesso_PRA_2004}
\begin{eqnarray}
    \textbf{d}=\begin{pmatrix}
\langle \hat{x}_A\rangle \\ \langle \hat{p}_A \rangle \\ \langle \hat{x}_B \rangle \\ \langle \hat{p}_B \rangle
\end{pmatrix}; ~~~
\Xi=\begin{pmatrix}
\mathcal{A} & \mathcal{B}\\
\mathcal{B} & \mathcal{C}
\end{pmatrix},
\label{eq:initial_state} 
\end{eqnarray}
where $\hat{x}_i, \hat{p}_i$ are respectively the position and momentum quadratures of the mode $i$, $\mathcal{A} = a I_2, \mathcal{C} = c I_2,$ and $\mathcal{B} = \text{diag}(b_1, b_2)$~\cite{Tserkis_PRA_2017} with $a,b_1, b_2,c$ being real numbers, and $I_2$ being the $2\times2$ identity matrix. While local symplectic operations preserve the entanglement between the sender and the receiver or change the conditional entropy, they can increase the energy of the state. Since the capacity is evaluated at fixed sender-side energy, such local symplectic operations may change the capacity and also impact its variation with respect to the negative conditional entropy, and therefore must be taken into account.
For a single sender–receiver pair, the most general local Gaussian symplectic operations consist of single-mode rotations, squeezing, and displacement~\cite{Adesso_OSID_2014}. Since rotations are passive transformations that do not change the energy of the state, and the parties are free to set the overall displacement to zero as a part of the encoding process, the relevant local operation reduces to squeezing on each mode. We, therefore, consider local squeezing operations of the form \(\hat S_i=\exp\!\big[\tfrac{1}{2}s_i(\hat a^2-\hat a^{\dagger 2})\big]\), characterized by real squeezing parameters \(s_1\) and \(s_2\), acting on the two
modes of the state $\rho_{AB}$, with $\hat{a}(\hat{a}^\dagger)$ representing the Bosonic annihilation (creation) operator. This alters the parameters of the covariance matrix to $\mathcal{A} = a ~\text{diag}(e^{2s_1},e^{-2s_1}), \mathcal{C} = c ~\text{diag}(e^{2s_2},e^{-2s_2}),$ and $\mathcal{B} = \text{diag}(b_1e^{s_1+s_2}, b_2e^{-(s_1+s_2)})$~\cite{Tserkis_PRA_2017}.
}
The quadrature expectation values comprising the displacement vector may be taken to be zero, since they can easily be manipulated using local displacement operations which do not affect the entanglement. Furthermore, it has been established that using local displacement operations to obtain a null displacement vector prior to the dense coding protocol helps to maximize the mutual information between the sender and the  receiver~\cite{Lee_PRA_2014}.

\textcolor{black}{In the Gaussian setting, when a single mode undergoes evolution while interacting with the environment, the decoherence dynamics can be well represented with the help of deterministic Gaussian completely positive (CP) maps~\cite{Serafini_2017}. Such maps are generally characterized by two operators $X$ and $Y$, represented by $2 \times 2$ matrices constituting the parameters modeling the noise.} The displacement vector and the covariance matrix of a two-mode state, $\rho_{AB}$, one of whose modes, say $A$, has undergone a noisy evolution, are given by

\begin{equation}
\textbf{d}'=\begin{pmatrix}
    X & 0 \\
    0 & I_2
\end{pmatrix} \begin{pmatrix}
 \langle \hat{x}_A\rangle \\ \langle \hat{p}_A \rangle \\ \langle \hat{x}_B \rangle \\ \langle \hat{p}_B \rangle
\end{pmatrix}; ~~
\Xi'=\begin{pmatrix}
 X \mathcal{A} X^{T} + Y & X \mathcal{B} \\
 \mathcal{B} X^T & \mathcal{C}
\end{pmatrix}
\label{eq:noisy_state}.
\end{equation}
One can recover the noiseless limit when $X = I_2$ and $Y = 0$. We now proceed to describe the noisy dense coding routine and derive its capacity.

\subsubsection*{Noise in state distribution }

Let us consider that the channels through which the two modes are distributed to the sender and the receiver are noisy and are respectively characterized by ~\cite{Serafini_2017}

\begin{align}
X_1=x_1I_2 \quad;\quad Y_1=y_1I_2 ,\\
X_2=x_2I_2 \quad;\quad Y_2=y_2I_2 ,
\label{eq:initial_noise}
\end{align}
where $x_1,y_1,x_2,y_2$ are the noise parameters. The displacement vector and the covariance matrix of the state after the noise has acted may be represented as
\begin{equation}
\textbf{d}_{\text{dist}}=\begin{pmatrix}
X_1&0\\
0&X_2
\end{pmatrix}\mathbf{d} ,
\label{eq:dist_d}
\end{equation}
\textcolor{black}{
\begin{equation}
\Xi_{\text{dist}}=\begin{pmatrix}
X_1 \mathcal{A} X_1^T + Y_1 &  X_1 \mathcal{B} X_2^T\\
( X_1 \mathcal{B} X_2^T)^T &  X_2 \mathcal{C} X_2^T + Y_2\\
\end{pmatrix} .
\label{eq:dist_cov}
\end{equation}}
Here we consider the initial displacement vector as $\mathbf{d} = (0, 0, 0, 0)^T$.

\subsubsection*{Noisy channel used to send the encoded mode }
 {\color{black}The sender can encode information in both the position} and momentum quadratures using a suitable displacement operation, $\hat{D}(\alpha) = \exp (\alpha \hat{a}^\dagger - \alpha^* \hat{a})$, where $\alpha = \alpha_x + \iota \alpha_p =  \frac{\langle\hat{x}\rangle + \iota \langle\hat{p}\rangle}{\sqrt{2}}$ is the displacement parameter. The message is encoded according to a Gaussian probability distribution of vanishing mean and standard deviation $\sigma$ given by

\begin{equation}
P(\alpha)=\frac{\exp[\frac{-(\alpha_x^2+\alpha_p^2)}{2\sigma^2}]}{2\pi\sigma^2},
\label{eq:P_alpha}
\end{equation}
while the displacement vector upon encoding assumes the form
\begin{equation}
    \mathbf{d}_{\text{en}}=\begin{pmatrix}
\sqrt2\alpha_x\\
\sqrt2\alpha_p\\
0\\
0
\end{pmatrix}
\label{eq:encoded_d}.
\end{equation}
Note that displacement operations do not affect the covariance matrix and thus, after encoding, $\Xi_{\text{en}} = \Xi_{\text{dist}}$.

Suppose the channel transmitting the encoded mode to the receiver is imperfect, which we represent using the noise operators, $X_3 = x_3 I_2$, and $Y_3 = y_3 I_2$. Such a noise would modify the encoded state as
\begin{equation}
    \mathbf{d}'_{\text{en}}=\begin{pmatrix}
X_3 &0\\
0 & I_2
\end{pmatrix}\mathbf{d}_{\text{en}} ,
\label{eq:d_en_noise}
\end{equation} and
\textcolor{black}{
\begin{equation}
  \Xi'_{\text{en}}  =\begin{pmatrix}
X_3 ( X_1 \mathcal{A} X_1^T + Y1) X_3^T + Y_3 &  X_3  X_1 \mathcal{B} X_2^T\\
( X_3  X_1 \mathcal{B} X_2^T)^T &  X_2 \mathcal{C} X_2^T+Y_2
\end{pmatrix}.
\label{eq:sigma_en_noise}
\end{equation}}
At this point, the receiver has both the modes of the two-mode state characterized by Eqs. \eqref{eq:d_en_noise} and \eqref{eq:sigma_en_noise}.

\subsubsection*{Imperfect double-homodyne measurement} 

\begin{figure}[ht]
\includegraphics[width=\linewidth, height = 5.5cm]{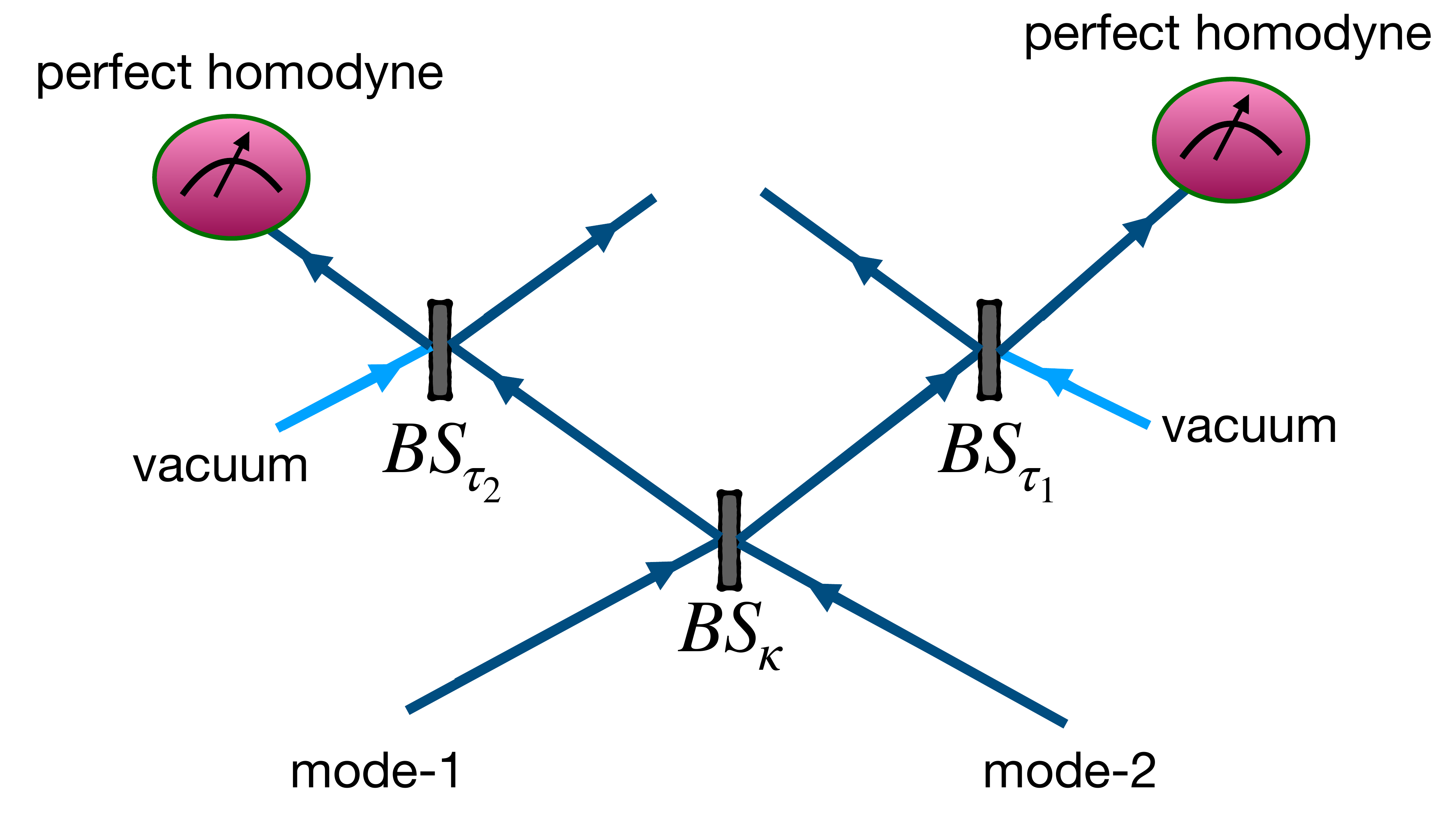}
\captionsetup{justification=Justified,singlelinecheck=false}
\caption{\textbf{Schematic diagram of the imperfect double-homodyne setup.} The two modes, mode-$1$ and mode-$2$, are initially impinged at an unbalanced beam splitter, $BS_\kappa$, of transmissivity $\kappa$. The subsequent output modes are mixed with vacuum at two fictitious beam splitters $BS_{\tau_i} (i = 1, 2)$ of transmissivity $\tau_i$. The transmitted modes from the fictitious beam splitters are subjected to perfect homodyne detection.}
\label{fig:imperfect_dh} 
\end{figure}

The decoding scheme comprising double-homodyne detection can also suffer from inaccuracies. Recall that a double-homodyne detection on two modes consists of mixing the modes at a $50/50$ beam splitter (BS), followed by homodyne detection of conjugate quadratures on the output modes. We can model the inexact decoding scheme in two ways \textbf{:} $(1)$ When instead of a $50/50$ beam splitter, one uses a BS with different transmissivity, $\kappa <1$, and $(2)$ when the imperfection comes from imperfect photon counting measurements comprising the double-homodyne setup. \textcolor{black}{We shall consider the second one in our work.} This can be modeled by two fictitious beam splitters with transmission coefficients $\tau_1$ and $\tau_2$, followed by perfect homodyne setups~\cite{Leonhardt_2000} as shown in Fig. \ref{fig:imperfect_dh}. Specifically, the fictitious beam splitter description involves mixing the two modes to be measured with vacuum at the two beam splitters, and the two corresponding transmitted modes thereafter undergo perfect homodyne measurement. Note that, when $\tau_1 = \tau_2$, and for the two-mode Gaussian state in Eq. \eqref{eq:d_en_noise}-\eqref{eq:sigma_en_noise}, the scheme is equivalent to mixing the two modes with vacuum first and then impinging the two transmitted modes on the BS with transmissivity $\kappa$. In our work, while computing the DCC, we consider $\tau_1 = \tau_2 = \tau$ and $\kappa = 1/2$, thereby quantifying the imperfection in the double-homodyne setup by a single parameter, $\tau < 1$. At $\tau = 1$, we recover the perfect decoding measurement.


The symplectic operator for a beam splitter operator with transmission coefficient $\tau$ is given as
\begin{equation}
    \textbf{BS}_\tau=\begin{pmatrix}
\sqrt\tau&0&\sqrt{1-\tau}&0\\
0&\sqrt\tau&0&\sqrt{1-\tau}\\
\sqrt{1-\tau}&0&-\sqrt\tau&0\\
0&\sqrt{1-\tau}&0&-\sqrt\tau
\end{pmatrix}.
\end{equation}
\textcolor{black}{Thus, the displacement vector and the covariance matrix of the transmitted state undergoing imperfect double-homodyne detection with $\tau_1=\tau_2=\tau$, and $\kappa = 1/2$, can be represented as}
\begin{equation}
    \mathbf{d}_{\text{imper}}= \begin{pmatrix}
\sqrt2x_3\alpha_x\sqrt\tau\\
\sqrt2x_3\alpha_p\sqrt\tau\\
0\\
0
\end{pmatrix},
\label{eq:d_final}
\end{equation}
\begin{equation}
   \text{and} ~ \Xi_{\text{imper}} = \begin{pmatrix}
    \mathcal{A}' & \mathcal{B}'\\
    \mathcal{B}' & \mathcal{C}'
    \end{pmatrix},
    \label{sigma_final}
\end{equation}
where 
\begin{eqnarray}
  \nonumber  \mathcal{A}'&=& \text{diag}(\mathcal{Y}_1,\mathcal{Y}_2),\\
   \nonumber \mathcal{B}'&=& x_1 x_2 x_3  \tau \times \text{diag} (b_1e^{s_1+s_2}, b_2e^{-(s_1+s_2)}) \\
\nonumber   \mathcal{C}'&=&\text{diag}(\mathcal{X}_1,\mathcal{X}_2),\\
 \nonumber   \text{where}~ \mathcal{Y}_i &=& \Big(1 + (-1 + {x_3}^{2} (a e^{(-1)^{(i+1)}2s_1} {x_1}^{2} + y_{1}) + y_{3})\tau \Big) \\
  \mathcal{X}_i &=& \Big(1 + (-1 + c e^{(-1)^{(i+1)}2s_2} {x_2}^2 + y_2) \tau \Big)
    \label{eq:final_cov_param}.
\end{eqnarray}
The receiver thereafter is considered to perform a perfect double-homodyne measurement by applying a $50/50$ beam splitter on the above state and then performing perfect homodyne measurements on both modes to infer the position quadrature of one mode, $\beta_x$, and the momentum quadrature of the other mode, $\beta_p$.
\textcolor{black}{The conditional probability distribution of the output variable, $\beta=\frac{\beta_x+\iota\beta_p}{\sqrt{2}}$, with respect to the input variable, $\alpha$, is given as }
\begin{equation}
    P(\beta|\alpha)=2\frac{\exp[\frac{-2(\beta_p - x_3 \alpha_p \sqrt{ \tau})^2}{\mathcal{G}_1}]}{\sqrt{\pi \mathcal{G}_1}}\frac{\exp[\frac{-2(\beta_x - x_3 \alpha_x \sqrt{\tau})^2}{\mathcal{G}_2}]}{\sqrt{\pi \mathcal{G}_2}},
\end{equation}
where \textcolor{black}{$\mathcal{G}_i = 2 + \Big(-2 + c e^{(-1)^{i}2s_2} x_2^2 + x_3 ((-1)^{i+1} 2 b_{(3-i)} e^{(-1)^i(s_1+s_2)} x_1 x_2 + a e^{(-1)^{i}2 s_1} x_1^2 x_3 + x_3 y_1) + y_2 + y_3 \Big) \tau$ for $i = 1,2$.}
We can easily calculate the unconditional output probability distribution, $P(\beta)$, and the mutual information, $\mathcal{I}$, corresponding to the noisy dense coding setup as
\begin{equation}
    P(\beta)=\int_{-\infty}^{\infty}P(\beta|\alpha)P(\alpha)\,d\alpha_xd\alpha_p,
\end{equation} and
\begin{eqnarray}
  \nonumber &&  \mathcal{I} = \int_{-\infty}^{\infty}P(\beta|\alpha)P(\alpha)\log(\frac{P(\beta|\alpha)}{P(\beta)})\,d\alpha_xd\alpha_pd\beta_xd\beta_p \\
 &=& \frac{1}{2}  \sum_{i = 1}^2 \log  \Big( 1 + \frac{4 x_3^2 \sigma^2 \tau}{\mathcal{G}_i}\Big)
      \label{eq:Inf}.
\end{eqnarray}
Note that the mutual information is a function, not only of the state parameters but also of the various imperfection variables. In the noiseless limit, i.e., $x_i = 1, y_i = 0, \tau = 1,s_1=0,s_2=0$, we have $\mathcal{I} = \log (1 + 2 e^{2r} \sigma^2)$ for a TMSV state of squeezing strength $r$.

\subsubsection*{Energy at the sender's side}
In order to calculate the dense coding capacity, we need to maximize the mutual information over the encoding parameter, $\sigma$, subject to a fixed number of photons, $\bar{N}_S$, at the sender's mode, i.e., $\textbf{C}=\underset{\sigma}{\max}~ \mathcal{I}$. Note that the photon number at the sender's mode refers to the total photon number after encoding, together with the effect of the noise and imperfections, and is given by
\begin{eqnarray}
\bar{N}_S&=&\int_{-\infty}^{\infty}N_\alpha P(\alpha)\,d\alpha_xd\alpha_p\,\\
\nonumber &=&\frac{(-1 + y_3 + x_3^2 (a x_1^2 \cosh{2s_1}+ y_1 + 4 \sigma^2)) \tau}{2},\\
\label{eq:Nmax}
\end{eqnarray}
where $N_\alpha$ quantifies the energy after the encoding process and can be calculated from $\Xi_{\text{imper}}$ and $\mathbf{d}_{\text{imper}}$ as
\begin{eqnarray}
  \nonumber  N_\alpha=\frac{1}{2}\,\tau &\Big(& -1 + y_3 + x_3^{2}\big(y_1 + 2(\alpha_p^{2} + \alpha_x^{2})\big)\\
&+& a\,x_1^{2} x_3^{2}\cosh2s_1 \Big).
\end{eqnarray}
\ms{The mutual information can be optimized in two ways - one when the noise is known to the parties, which we refer to as the \textit{adaptive scheme} while in the other case, the users are oblivious to the presence of noise, which we call the \textit{non-adaptive scheme}. Note that, by adaptive, we mean that the participants can adapt or tailor the protocol to counter the destructive impact of noise. We do not consider any feedback-based adaptive mechanism in this work.} Below, we derive the dense coding capacity for both these scenarios. 

\subsubsection*{Adaptive and non-adaptive dense coding capacity} 
\textcolor{black}{Unlike finite-dimensional systems, where the Holevo quantity can be maximized to derive the channel capacity of Pauli channels under the adaptive scheme~\cite{Laurenza_PRR_2020_channel-capacity}, due to the energy constraints in CV setups, we must optimize the mutual information between the sender and the receiver.} Since the noise is known to the involved parties, we can calculate the encoding parameter $\sigma$ from Eq. (\ref{eq:Nmax}) as
 \begin{equation}
     \sigma^{\text{ad}} = \frac{\sqrt{2 \bar{N}_S + (1 - x_3^2 (a \cosh{2s_1} x_1^2 + y_1) - y_3) \tau}}{
 2 x_3 \sqrt{\tau}}.
 \label{eq:sigma}
 \end{equation}
 Substituting the above expression in Eq. (\ref{eq:Inf}), we can maximize the mutual information with respect to input parameter and obtain adaptive capacity \(\tilde{\mathbf{C}}_{\text{ad}}=\mathcal{I} (\bar{N}_S,\tau,x_1,y_1,x_2,y_2,x_3,y_3)\). \ms{Note that the DCC obtained in this form corresponds to the entanglement-assisted capacity of transmitting classical information and does not correspond to the channel capacity in the general sense.}
 
 For a TMSV state with $a=c=\cosh2r$, and $b_1 = -b_2 =\sinh2r$, $s_1=s_2=0$, \textcolor{black}{ which possesses the highest DCC in the noiseless case (see Appendix \ref{app:tmsv_cap}), the optimal squeezing strength can be obtained by substituting $\sigma_{\mathrm{ad}}$ in the expression for the mutual information and maximizing it over the parameters, as}
 
 \begin{widetext}
 \begin{eqnarray}
   \nonumber (r^{\text{opt}}_{\text{ad}})_{\text{TMSV}} &=& \frac{1}{2} \log \Big( \frac{1}{2 \bar{N}_{S} (x_2 - 
         x_1 x_3)^2 - x_2^2 (-1 + x_3^2 y_1 + y_3) \tau + 
      2 x_1 x_2 x_3 (-1 + x_3^2 y_1 + y_3) \tau + 
      x_1^2 x_3^2 (2 + (-1 + y_2) \tau)} \times \\
    \nonumber && [ -2 x_1^3 x_2 x_3^3 \tau + 
     \frac{1}{2} \{16 x_1^6 x_2^2 x_3^6 \tau^2 + 
          4 (2 \bar{N}_S (x_2 + x_1 x_3)^2 - 
             x_2^2 (-1 + x_3^2 y_1 + y_3) \tau - 
             2 x_1 x_2 x_3 (-1 + x_3^2 y_1 + y_3) \tau \\ \nonumber && + 
             x_1^2 x_3^2 (2 + (-1 + y_2) \tau)) (2 \bar{N}_S (x_2 - 
                x_1 x_3)^2 - x_2^2 (-1 + x_3^2 y_1 + y_3) \tau + 
             2 x_1 x_2 x_3 (-1 + x_3^2 y_1 + y_3) \tau \\ &&  + 
             x_1^2 x_3^2 (2 + (-1 + y_2) \tau))\}^{1/2} ] \Big).
             \label{eq:r_opt}
 \end{eqnarray}
  \end{widetext}
Using the above form of $(r^{\text{opt}}_{\text{ad}})_{\text{TMSV}}$, the behavior of the DCC for the TMSV state in the adaptive situation can be analyzed with respect to specific noise models and imperfection, $\{x_i,y_i\}$ $(i=1, 2, 3)$ and $\tau$ (see Sec. \ref{subsec:TMSV_noise_case}).
 
If the parties are unaware of the presence of system imperfections, the DCC can be estimated using the optimal input distribution obtained for the noiseless case.
In practice, this amounts to substituting
$\sigma=\sqrt{2\bar N_S+1-a\cosh(2s_1)}/2$ into Eq.~(\ref{eq:Inf}) and maximizing the noiseless mutual information over the input state parameters $a,b_1,b_2,c,s_1,$ and $s_2$, at fixed noiseless energy, and substituting the corresponding maximum values in the expression for the noisy mutual information, yielding the non-adaptive capacity
$\tilde{\mathbf{C}}_{\text{non-ad}}$.

 In particular, for the TMSV state the optimal non-adaptive squeezing parameter is
 \begin{eqnarray}
     (r^{\text{opt}}_{\text{non-ad}})_{\text{TMSV}}=\frac{1}{2} \log(1+2\bar{N}_S),
     \label{eq:ropt-na}
 \end{eqnarray}
where $\bar{N}_S = \sinh^2 r + 2 \sigma^2$ which leads to the non-adaptive DCC.

\textbf{Note $\mathbf{1}$.} In actual experiments, a TMSV state of fixed squeezing strength $r$ is provided. Knowing the expression for $(r^{\text{opt}}_{\text{ad/non-ad}})_{\text{TMSV}}$, one can calculate $\bar{N}_S$ and hence the optimal encoding distribution, $\sigma$, to obtain the DCC.

\begin{figure*}[ht]
\includegraphics[width=\linewidth]{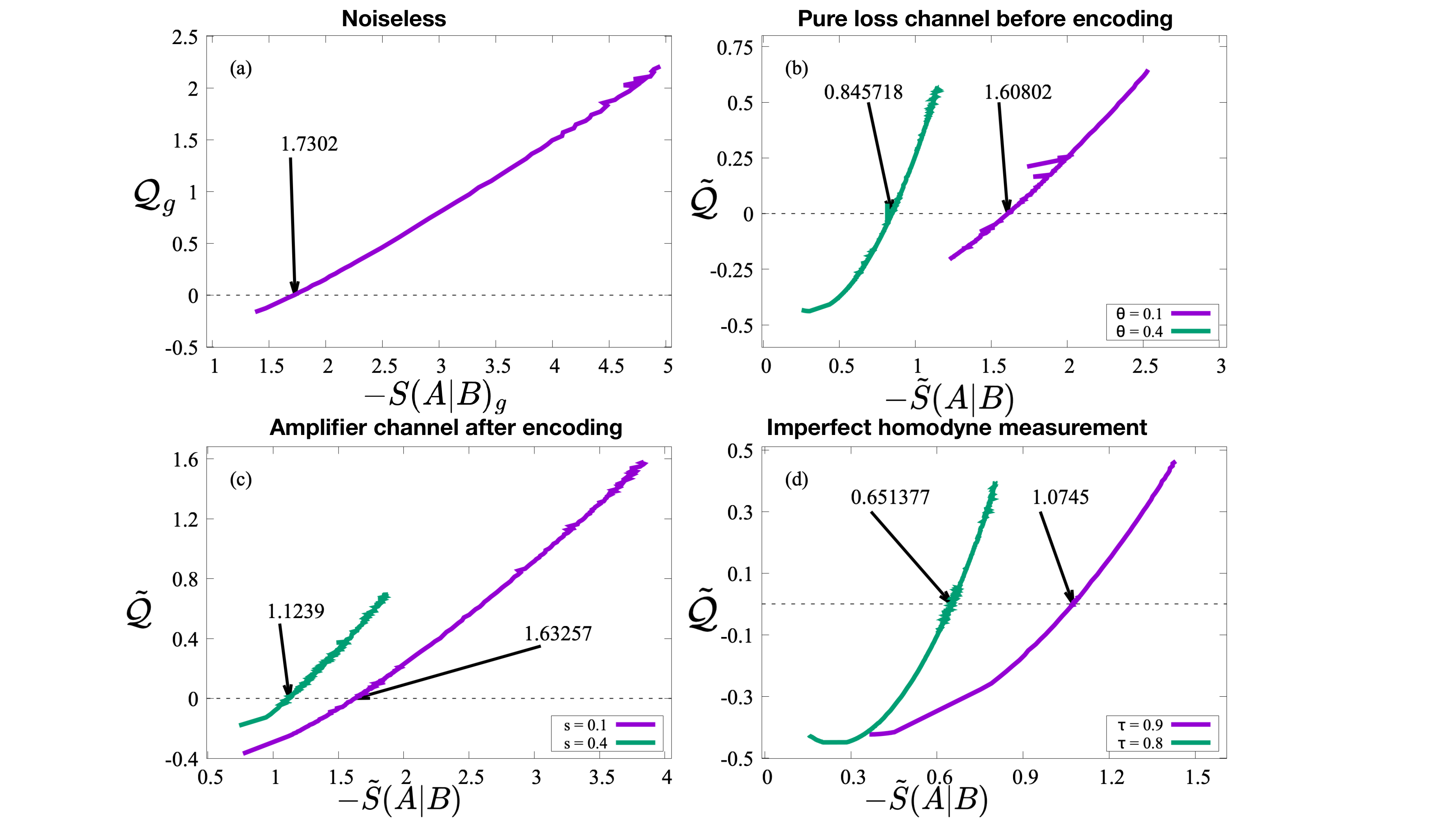}
\captionsetup{justification=Justified,singlelinecheck=false}
\caption{\textcolor{black}{\textbf{Quantum advantage and negative conditional entropy for the state described in Eq.~(\ref{eq:initial_state}) both with and without imperfect setups.} (a) The variation of the quantum advantage, \(\tilde{\mathcal{Q}}_g\) (ordinate), as a function of the negative conditional entropy, \(-\tilde{S}(A|B)_g\) (abscissa), for the general state described in Eq.~(\ref{eq:initial_state}) under ideal conditions. Panels (b) and (c) represent the same for pure-loss channel before encoding and quantum-limited amplifier noise after encoding for noise strengths $s (\theta) =0.1, 0.4$ in purple (dark) and green (light) respectively. In panel (d), the effect of the imperfect homodyne detection scheme is illustrated for $\tau = 0.9 ~\text{and}~ 0.8$ in purple (dark) and green (light) respectively.  Both axes are in bits. }}
\label{fig:general_all} 
\end{figure*}

\section{Analysis of dense coding capacity under paradigmatic noise models}
\label{sec:dcc_analysis_exemplarychannel}
Let us now present some typical noise models which we consider to act on the shared resource before and after the encoding process.
\begin{itemize}
    \item \textbf{Amplifier channels: } When the state interacts with a thermal environment of mean photon number, $\bar{N}$, through a two-mode squeezing operation, the resulting noisy channel is referred to as the \textit{amplifier channel}~\cite{Serafini_2017}. The deterministic Gaussian CP map corresponding to such a channel is characterized by
    \begin{equation}
        X = \cosh s I_2; ~~ Y = n_{\text{th}} \sinh^2 s I_2,
        \label{eq:amp_noise}
    \end{equation}
    where $0 \leq s < \infty$ and $n_{\text{th}} = (2 \bar{N} + 1) \geq 1$. Such a channel represents the enhancement in the amplitude of the input state. When $n_{\text{th}} = 1$, we obtain the \textit{quantum-limited amplifier channel}, which we will consider in our calculations. 

    \item \textbf{Attenuator channels: } Attenuator channels describe the mixing of the input state with a thermal environment with $n_{\mathrm{th}} = (2 \bar{N} + 1) \geq 1$ at a beam splitter~\cite{Serafini_2017}. Such channels denote the reduction of the first moments of the affected state and are defined by 
    \begin{equation}
        X = \cos \theta I_2; ~~ Y = n_{\text{th}} \sin^2 \theta I_2,
        \label{eq:att_noise}
    \end{equation}
    with $\theta \in [0, 2 \pi]$. We deal with a special class of attenuator channels for which $n_{\text{th}} = 1$, known as the \textit{pure-loss channel}.

    \textcolor{black}{Note that, the beamsplitter transmissivity, say $t$, is related to the beamsplitter angle, $\theta$, as $t = \cos^2 \theta$~\cite{Serafini_2017}. In this study, we choose to illustrate the variation of the dense coding capacity against $\theta$ (and not with respect to $t$ or the transmission distance) since it is what characterizes the noise, and we are more interested in studying the direct impact of the same on the dense coding protocol. }

    \item \textbf{Environmental noise: } \textcolor{black}{When the constituent modes interact with the environment during transmission, the} system-environment interaction, for a single mode, may be characterized by the Gorini-Kossakowski-Sudarshan–Lindblad master equation~\cite{Gorini_JMP_1976, Lindblad_CMP_1976} under the paradigm of open system dynamics as

\begin{equation}
    \frac{d \rho}{d t} = \gamma (\bar{N} + 1) D_{\rho}[\hat{a}] + \gamma \bar{N} D_{\rho}[\hat{a}^{\dagger}],
    \label{eq:master_eqn}
\end{equation}
where $\rho$ is the density matrix of a single mode system which admits the creation(annihilation) operator $\hat{a}^{\dagger}(\hat{a})$ and the dissipator $D_{\rho}[\hat{o}] = \hat{o} \rho \hat{o}^{\dagger} - \frac{1}{2}\{\rho, \hat{o}^{\dagger} \hat{o}\}$ with $\gamma$ denoting the coupling strength between the system and the environment. Such an interaction eventually drives the system to a thermal state of mean photon number $\bar{N}$ in the long-time limit. For Gaussian systems, such environmental noise may be modeled by~\cite{Alves_arXiv_2024}

\begin{equation}
    X = \exp(-\frac{\gamma t}{2}) I ~~ \text{,}~~ Y = (2 \bar{N} + 1)(1 - e^{-\gamma t}) I.
    \label{eq:env_noise}
\end{equation}
where $t$ denotes the time for which the interaction takes place.

\end{itemize}

{\color{black}

\subsection{Noisy dense coding with generic Gaussian states: role of conditional entropy under imperfections}
\label{subsec:general_states} 

\begin{figure*}[ht]
\includegraphics[width=\linewidth, height=10cm]{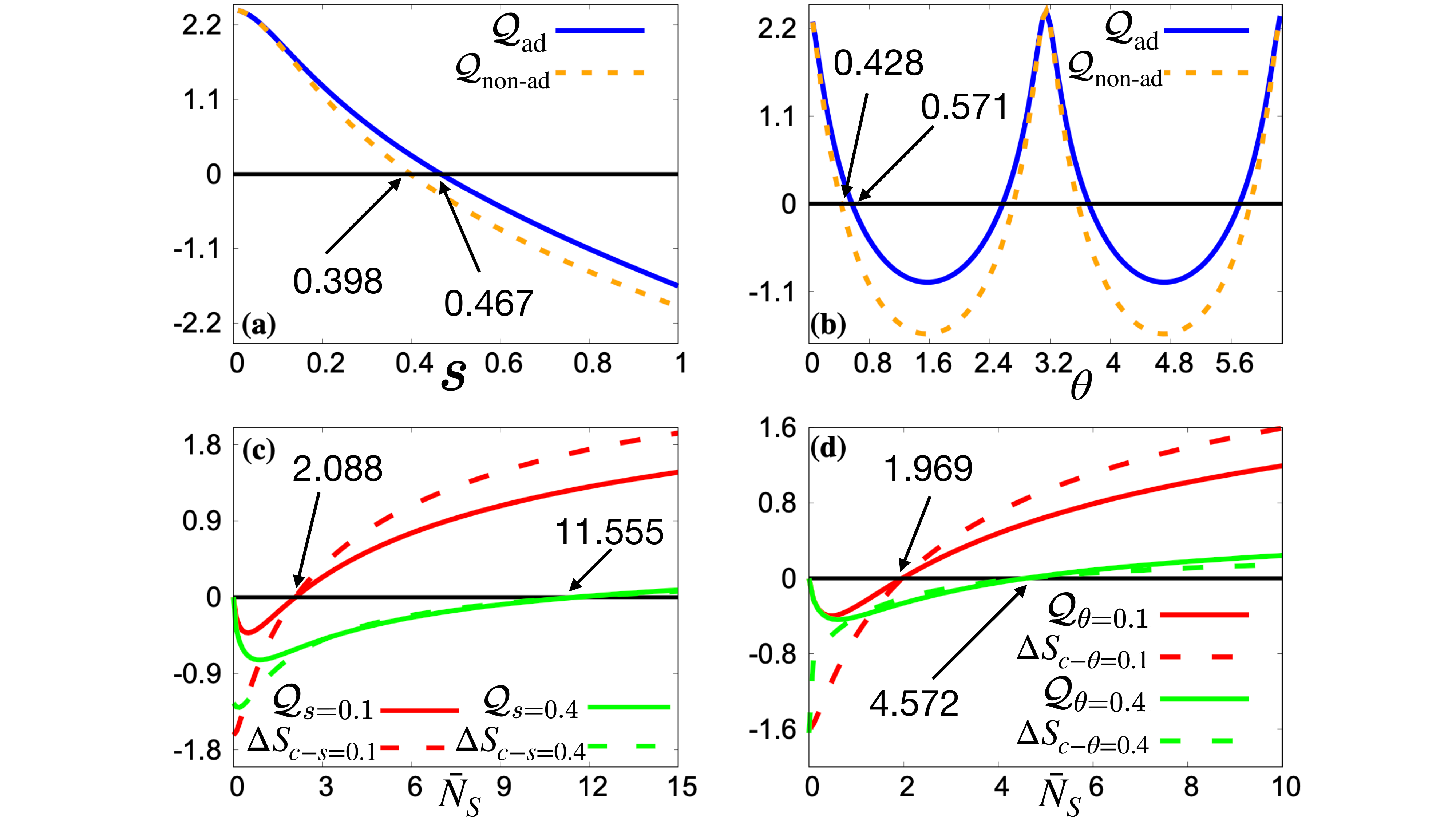}
\captionsetup{justification=Justified,singlelinecheck=false}
\caption{\textbf{Quantum advantage and negative conditional entropy for the TMSV state affected by noise during distribution.} (Upper panel) The behavior of the adaptive quantum advantage, $\mathcal{Q}_{\text{ad}}$ (light orange solid line), the non-adaptive quantum advantage (light orange dashed line), along the ordinate against the noise parameter. (a) the quantum-limited amplifier with noise parameter $s$, and (b) the pure loss channel with noise strength $\theta$. The sender-side photon number is fixed to $\bar{N}_S = 30$. (Lower panel) The variation of the quantum advantage, $\mathcal{Q}$ (solid line), and the conditional entropy difference, $\Delta S_c$ (dashed line), along the ordinate is demonstrated against the sender-side energy, $\bar{N}_S$ (abscissa) for the quantum-limited amplifier with noise strength $s = 0.1$ (dark red line) and $s = 0.4$ (light green line) in (c), and for the pure-loss channel with $\theta = 0.1$ (dark red line) and $\theta = 0.4$ (light green line) in (d). The horizontal axis is dimensionless whereas the vertical axis is in bits.}
\label{fig:dist-noise} 
\end{figure*}

\ms{Let us investigate whether the conditional entropy can be used to predict quantum advantage even when the Gaussian dense coding protocol is affected by noise. We have already shown that, for a noiseless TMSV state, the negative conditional entropy can emulate the quantum advantage. We first extend this analysis to investigate whether negative conditional entropy continues to play the same role for more general quantum states in the ideal scenario.} Accordingly, for the general class of local symplectic rotated states obtained from Eq.~(\ref{eq:initial_state}), the quantum advantage of dense coding is quantified as
\(\mathcal{Q}_g = \tilde{\mathbf{C}}_g - \mathbf{C}_{\text{cl}}\), where $\tilde{\mathbf{C}}_g = \tilde{\mathbf{C}}_{\mathrm{ad}}(x_1=x_2=x_3=1, y_1=y_2=y_3=0, \tau = 1)$. The quantum advantage $\mathcal{Q}_g$ becomes non-negative only when the sender’s energy exceeds the threshold value  \(\bar{N}_{S_g-\text{th}} = 1.8849\),  which corresponds to a threshold negative conditional entropy of  \(-S(A|B)_{g-\text{th}} = 1.73023\).  \ms{Defining \(\Delta S_{c_g} = -S(A|B)_g + S(A|B)_{g-\text{th}}\), 
we observe from Fig.~\ref{fig:general_all} (a) that beyond this threshold, 
both \(\Delta S_{c_g}\) and \(\mathcal{Q}_g\) increase simultaneously, 
indicating a direct correspondence between the negative conditional entropy and the quantum advantage.}

We now turn to the noisy scenario.
As a first step, we consider the adaptive dense-coding protocol and investigate whether the general class of states continues to exhibit \(\tilde{\mathcal{Q}} = \tilde{\mathbf{C}}_{\text{ad}} - \mathbf{C}_{cl} >0\), even under noise, beyond a threshold value of negative conditional entropy.
\ms{For all considered noise models, the quantum advantage, \(\tilde{\mathcal{Q}}\), and the quantity \(\Delta \tilde{S}_c= -\tilde{S}(A|B) + \tilde{S}(A|B)_{\text{th}}\) behave in a similar fashion with respect to \({{\bar{N}}_S} \) beyond the threshold values \(-\tilde{S}(A|B)_{\text{th}} = 0.846 ~\text{and}~ 1.608\) for the noise parameters \(\theta = 0.4 ~\text{and}~ 0.1\) respectively.} Crucially, beyond  \(-\tilde{S}(A|B)_{\text{th}}\), the quantum advantage exhibits strictly positive values (see Fig.~\ref{fig:general_all} (b)).
Similarly, for post-encoding quantum limited amplifier noise, quantum advantage exists beyond \(-\tilde{S}(A|B)_{\text{th}}= 1.124 ~\text{and} ~1.633\) at $s = 0.4~\text{and}~ 0.1$ as shown in Fig.~\ref{fig:general_all} (c). Furthermore, even under imperfect decoding schemes, the exact same relationship between the negative conditional entropy and the quantum advantage is evident from Fig.~\ref{fig:general_all}(d).
Therefore, even for a state with local symplectic operations, the protocol maintains its quantum advantage - it is non-vanishing throughout the persistence of positive values of \(\Delta \tilde{S}_c\). \ms{A similar analysis is carried out in Appendix~\ref{app:nonadaptive_dc} for the non-adaptive noisy dense coding scenarios, confirming that even in the absence of adaptive optimization, negative conditional entropy remains a sufficient indicator of quantum advantage in Gaussian CV dense coding.}


\subsection{The noisy TMSV state as a special instance}
\label{subsec:TMSV_noise_case}

The TMSV state is one of the most fundamental and widely used resources in continuous-variable quantum information. For a fixed energy, it maximizes entanglement among all two-mode Gaussian states~\cite{Wolf_PRL_2006_extremal_Gaussian-states, Adesso_JPA_2007}, playing a role analogous to that of a Bell state in discrete-variable systems. Moreover, the TMSV state is experimentally accessible with current technologies~\cite{Li_PRR_2021_TMSV-experiment, Song_PRA_2023_TMSV-experiment}, and serves as the cornerstone for quantum advantage in CV teleportation~\cite{Furusawa_Science_1998_teleportation}, key distribution~\cite{Gehring_NatComm_2015_qkd, Jacobsen_NPJQI_2018_qkd}, and quantum illumination~\cite{Tan_PRL_2008_illumination}. Motivated by its central importance, we now explicitly analyze the performance of the TMSV state under the action of various channel noise models, including measurement imperfections.

\subsubsection{Influence of amplifier and pure-loss channel on shared TMSV state}
\label{subsubsec:distribution_noise}

\begin{figure*}[ht]
\includegraphics[width=\linewidth]{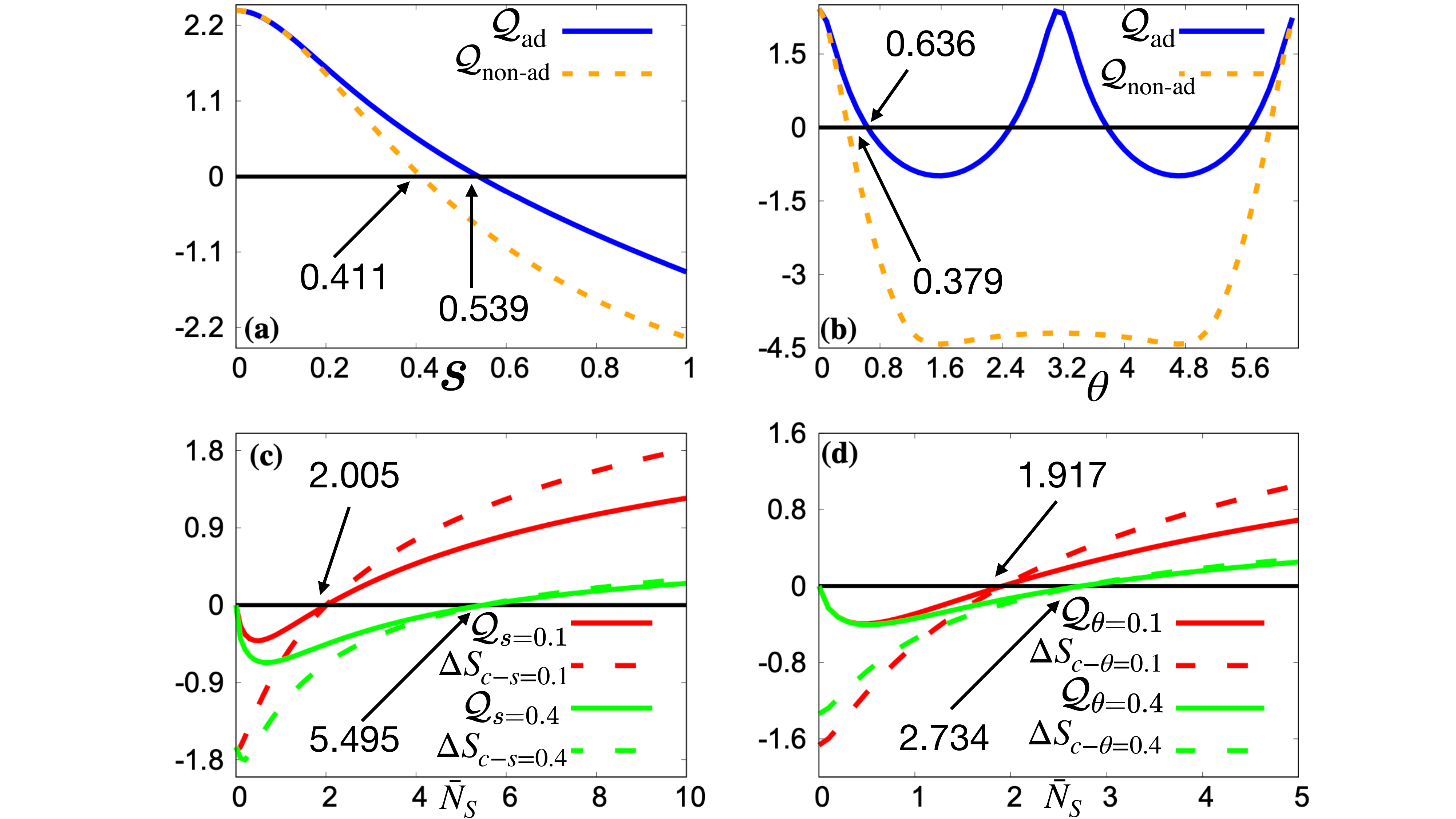}
\captionsetup{justification=Justified,singlelinecheck=false}
\caption{\textbf{\textcolor{black}{Quantum advantage and negative conditional entropy for the TMSV state affected by noise after encoding.}} All specifications are the same as in Fig. \ref{fig:dist-noise}}
\label{fig:encode_noise} 
\end{figure*}

Let us first assume that the channels that are used to share the resource state between the sender and the receiver are noisy, i.e., the quantum-limited amplifier or the pure-loss channel act on both the modes of shared state, before encoding. This translates to assuming $x_1 = x_2 = \cosh s, y_1 = y_2 = \sinh^2 s$ for the former case while $x_1 = x_2 = \cos \theta, y_1 = y_2 = \sin^2 \theta$ for the latter.

In the case of the quantum-limited amplifier channel, the adaptive capacity decreases with the increase of $s$ and it exceeds the corresponding classical threshold, e.g., when $s \lesssim 0.467$ for a fixed sender-side energy of $\bar{N}_S = 30$. Specifically, in that range of the noise strength, we obtain $\mathcal{Q} = \tilde{\mathbf{C}}_{\text{ad}} - \mathbf{C}_{\text{cl}} > 0$ (see Fig. \ref{fig:dist-noise} (a)). In a similar setting of $\bar{N}_S = 30$, the situation worsens in the non-adaptive case, whence $\mathcal{Q} > 0$ only in a more limited noise range, $s \in (0, 0.398)$. This clearly indicates that prior knowledge of the noise is beneficial in obtaining non-classical capacity against a higher amount of noise as compared to the non-adaptive case.

On the other hand, for the pure-loss channel, the quantum advantage oscillates between positive and negative values due to the periodic nature of the noise operator, both in the adaptive and the non-adaptive case (see Fig. \ref{fig:dist-noise} (b)). In fact, at $\theta = p \pi$ with $p \in \text{integers}$, the quantum advantage attains its noiseless value. Once again, the adaptive strategy proves superior to the non-adaptive case since the quantum advantage persists over a larger range of the noise strength as shown in Fig. \ref{fig:dist-noise} (b). \textcolor{black}{For example, the region where the quantum advantage vanishes is given by $0.571 \leq \theta \leq 2.57$ (adaptive) and $0.428 \leq \theta \leq 2.713$ (non-adaptive) at $\bar{N}_S = 30$.}

\textcolor{black}{The variation of the adaptive quantum advantage and the conditional entropy difference are observed to be in one-to-one correspondence, akin to the noiseless TMSV scenario.} For the quantum-limited amplifier, the threshold energy for quantum advantage is given by $\bar{N}_{S-\text{th}} = 2.088, 11.555$ for $s = 0.1, 0.4$ respectively. We observe that beyond these threshold energies, a higher value of $\Delta S_c = -S(A|B)(\bar{N}_S, s=0.1, 0.4) + S(A|B)(\bar{N}_{S-\text{th}}, s = 0.1,0.4)$ indicates a greater quantum advantage (see Fig. \ref{fig:dist-noise}(c)). A similar behavior emerges for the pure-loss channel (see Fig. \ref{fig:dist-noise}(d)), whence $\bar{N}_{S-\text{th}} = 1.969, 4.572$ for $\theta = 0.1, 0.4$ respectively. 
\ms{Therefore, even though the analytical form of the conditional entropy is too complicated to investigate its behavior, numerical studies corroborate the fact that it is a precursor to the success of the CV dense coding protocol using Gaussian states, even when the system is affected by noise.} It is also notable that the threshold sender-side energy increases with the increase in noise strength, which is an expected result due to the destructive influence of noise. 


\subsubsection{Noise after encoding}
\label{subsubsec:encoding_noise}

\begin{figure*}[ht]
\includegraphics[width=1\linewidth]{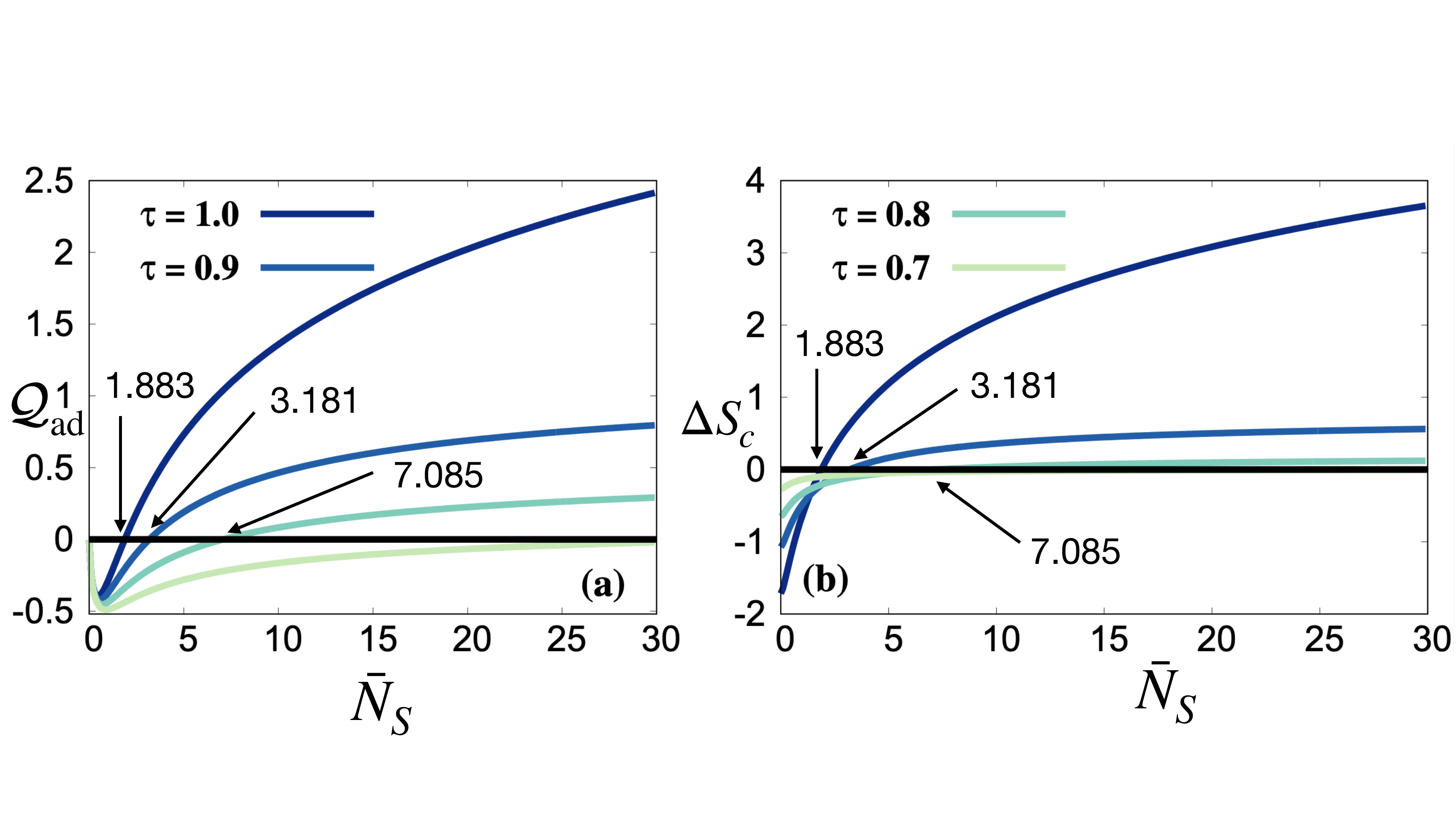}
\captionsetup{justification=Justified,singlelinecheck=false}
\caption{\textbf{Quantum advantage and negative conditional entropy after imperfect homodyne detection for the TMSV state.} (a)  $\mathcal{Q}_{\text{ad}}$ (ordinate) and (b) $ \Delta S_c$ (ordinate)  against  $\bar{N}_S$ (abscissa). From dark to light lines, the imperfections are $\tau = 1$ (perfect), $\tau = 0.9$, $\tau = 0.8$, and $\tau = 0.7$ respectively.  The y-axis is in bits, whereas the x-axis is dimensionless.}
\label{fig:imper-homo} 
\end{figure*}

Let us now consider the scenario when the encoded mode is sent to the receiver through a noisy quantum channel, although the sharing of the state and the decoding process are perfect, i.e., when $x_1 = x_2 = 1, y_1 = y_2 = 0$ and $\tau = 1$. Akin to the previous discussion, we set $x_3 = \cosh s,~ y_3 = \sinh^2 s$  for the quantum-limited amplifier whereas for the pure-loss channel, $x_3 = \cos \theta,~ y_3 = \sin^2 \theta$.
The behavior of the quantum advantage, $\mathcal{Q}$, is qualitatively similar to the case when noise acts during the state distribution (see Figs. \ref{fig:encode_noise} (a) and (b)). \textcolor{black}{Specifically, the threshold noise parameters, when the quantum advantage vanishes, are given by $s_{\text{th-ad}} = 0.539$, and $\theta_{\text{th-ad}}=0.636$ and,  while $s_{\text{th-non-ad}} = 0.411$, and $ \theta_{\text{th-non-ad}}=0.379$}. 
{\color{black} Furthermore, in the pure-loss channel, the revival of the quantum advantage occurs roughly twice as often in the adaptive scenario as in the non-adaptive one. However, this is not a strict quantitative relation but a numerical coincidence specific to the chosen parameter regime in Fig. \ref{fig:encode_noise}(b). A possible argument for this observation may lie in the fact that the deterministic Gaussian CP map modeling comprises terms periodic in the noise parameter, like $\sin\theta$ and $\cos\theta$. Therefore, if the parties have prior knowledge of the noise, the adaptive scheme allows them to retain the periodic behavior of the capacity to a greater extent, even though noise after the encoding process diminishes the non-adaptive capacity much more drastically than in the case of noise during state distribution.}
We also notice that noise on the encoded mode furnishes a quantum advantage over a larger range of noise strength as compared to the case when noise acts on the shared state. This higher robustness to noise may be attributed to the fact that only the encoded mode is affected by the noise while the other mode remains untouched. The quantum advantage, when investigated against the sender-side energy, follows the exact same behavior as $\Delta S_c$ beyond the threshold energy for both considered noise models (see Figs. \ref{fig:encode_noise} (c) and (d)), \ms{thereby allowing us to identify it using the conditional entropy in CV dense coding using Gaussian resource states.}

\subsubsection{Imperfect homodyne detection}
\label{subsubsec:imper_homo}

\begin{figure*}[ht]
\includegraphics[width=\linewidth]{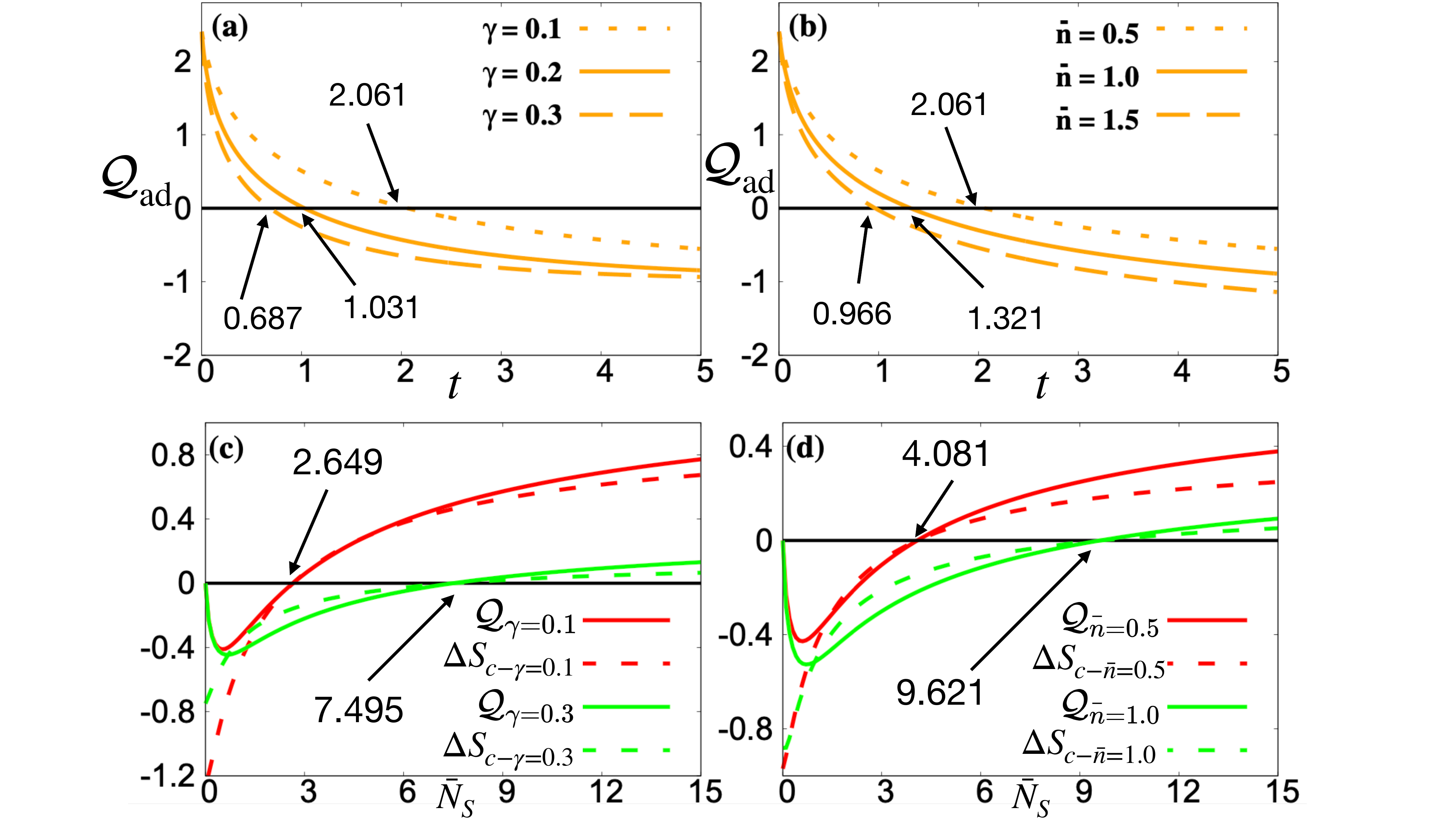}
\captionsetup{justification=Justified,singlelinecheck=false}
\caption{\textbf{Quantum advantage and negative conditional entropy for the TMSV state upon interaction with the environment.} (Upper panel) The  adaptive quantum advantage, $\mathcal{Q}_{\text{ad}}$ (ordinate), with respect to the interaction time, $t$ (abscissa) when (a) $\bar{n} = 0.5$ at the coupling strength $\gamma = 0.1$ (dotted line), $\gamma = 0.2$ (solid line), $\gamma = 0.3$ (dashed line), and (b) $\gamma = 0.1$ with mean environment photon number $\bar{n} = 0.5$ (dotted line), $\bar{n} = 1.0$ (solid line), $\bar{n} = 1.5$ (dashed line). In both situations, $\bar{N}_S = 30$. The y-axis is in bits whereas the x-axis is in seconds. (Lower panel) The variation of the quantum advantage, $\mathcal{Q}$ (solid line) and the conditional entropy difference (dashed line) (ordinate) against the sender-side energy, $\bar{N}_S$ (abscissa) when (c) $\bar{n} = 0.5$ is fixed and  $\gamma = 0.1$ (dark red line) and $\gamma = 0.3$ (light green line), (d) $\gamma = 0.1$ while $\bar{n} = 0.5$ (dark red line) and $\bar{n} = 1.0$ (light green line). The interaction time is taken to be $t = 0.5$ in both the cases. The environmental mean photon number is $\bar{N} = (\bar{n} - 1/2)/2$. The abscissa is dimensionless whereas the ordinate is in bits.}
\label{fig:env_noise} 
\end{figure*}

To visualize the effects of imperfection in the measurement setup on the DCC, all other kinds of noise, during distribution and after encoding, are considered to be absent, and hence the quantum advantage is a function of only the sender-side energy, $\bar{N}_S$ and $\tau$ of the BS before perfect double-homodyne measurement.

It is observed that as $\tau$ decreases (the double-homodyne measurement becomes more faulty), the threshold sender's mode photon number, at which quantum advantage is obtained, also increases (see Fig. \ref{fig:imper-homo}(a)). For instance, we find $\bar{N}_{S-\text{th}} = 1.883, 3.181, 7.085$ when $\tau = 1, 0.9, 0.8$ respectively. 
Our numerical calculations suggest that when the maximum sender-side energy is constrained to $\bar{N}_{S-\max} = 30$, the threshold imperfection reads as $\tau_{\text{th}} = 1/\sqrt{2}$ for the adaptive scheme and $\tau_{\text{th}} \approx 0.85$ for the non-adaptive case. Therefore, we can see that the dense coding protocol is not robust against very high imperfections in the decoding mechanism for moderate energy at the sender's mode.

Further, comparing Figs. \ref{fig:imper-homo} (a) and (b), it is evident that as long as the energy at the sender's side is above a certain threshold, \ms{the quantum advantage is indicated by the negative conditional entropy.}


\subsubsection{Noisy channels used in  state distribution and after encoding}
\label{subsubsec:dcc_env}

Let us now consider a situation in which the entire DC protocol is implemented in a noisy environment as described by Eq. (\ref{eq:master_eqn}). Such noise would impact the resource both during distribution between the participants and also after the encoding process. Therefore, while calculating the adaptive capacity, we can set $x_1 = x_2 = x_3 = \exp(-\frac{\gamma t}{2})$ and $y_1 = y_2 = y_3 = (2 \bar{N} + 1)(1 - e^{-\gamma t})$. Note that, since we have already established that the capacity is greater in the adaptive scheme, we restrict our analysis here to the same.

With the variation of the interaction time, $t$, for $\bar{n} = 0.5$ and different coupling strengths, we observe that the quantum advantage decreases monotonically with $t$, vanishing much faster as the coupling with the environment grows. Specifically, we have $t_{\text{th}} \approx 2.061, 1.031, 0.687$ for $\gamma = 0.1, 0.2, 0.3$ respectively (see Fig. \ref{fig:env_noise} (a)), where we define $t_{\text{th}}$ as the threshold time for obtaining  quantum advantage. This can be explained by the fact that a higher interaction rate with the environment drives the involved state towards a thermal state much faster, thereby causing the quantum advantage to persist only for short periods of time.

To analyze the effect of the mean photon number of the environment, $\bar{n}$, on the success of the protocol, we focus on the behavior of the quantum advantage at various values of $\bar{n}$ at $\gamma = 0.1$. Fig. \ref{fig:env_noise} (b) depicts that the $\mathcal{Q}_{\text{ad}}$ vanishes faster for higher $\bar{n}$, i.e., $t_{\text{th}} \approx 2.061, 1.321, 0.966$ for $\bar{n} = 0.5, 1, 1.5$. This may be attributed to the fact that a higher mean environmental photon number implies that the system interacts with an environment of higher temperature, leading to faster degradation of entanglement and hence a quicker decline in capacity.

Interestingly, the rate of decrease of the quantum advantage with $\bar{n}$ is slower than that with $\gamma$ (comparing Figs. \ref{fig:env_noise} (a) and (b)). To analyze this observation, let us define the change in threshold interaction time by $\delta t_{\text{th}} = t_{\text{th}}^i - t_{\text{th}}^f$, the change in coupling strength by $\delta \gamma = \gamma^f - \gamma^i$, and the change in the mean background environmental photon number by $\delta \bar{n} = \bar{n}^f - \bar{n}^i$, where the superscripts, $i, f$ stand for initial and final instants respectively. Quantitatively then, we find that $\delta t_{\text{th}} \approx 1.03$ when $\delta \gamma \approx 0.1$ while $\delta t_{\text{th}} \approx 0.74$ when $\delta \bar{n} \approx 0.5$. This means that the quantum advantage persists for a much shorter interaction time when the coupling strength changes even by a small value as compared to the change in the thermal photon number. \textcolor{black}{Therefore, we can conclude that the coupling strength with the environment affects the protocol much more drastically than the environment's thermal photon number.}

Even though open system dynamics is different from the case of noisy transmission through imperfect channels, \ms{we can still argue that the negative conditional entropy difference, $\Delta S_c$, can signal quantum advantage in this scenario.} Specifically, for a fixed interaction time, $t = 0.5$, we study the variation of the quantum advantage, and the conditional entropy difference at different coupling strengths (Fig. \ref{fig:env_noise}(c)) and different mean thermal photon number (Fig. \ref{fig:env_noise}(d)) against the sender-side energy. \ms{It is evident that beyond the threshold energy, both $\mathcal{Q}$ and $\Delta S_c$ are monotonic with each other, thereby highlighting the role of conditional entropy in the noisy dense coding protocol using Gaussian states.}

\section{TMSV superiority versus other pure Gaussian states}
\label{section:tmsv_other_Gaussian}

So far, we have analysed the performance of generic Gaussian states, with special emphasis on the TMSV state, in the dense coding protocol affected by decoherence and imperfections. Of course, compared to generic {\it mixed} Gaussian states, the TMSV emerges superior due to its high entanglement content. However, is it the most resourceful Gaussian state when pitted against other pure Gaussian states? In this section, we try to answer this question. We choose two situations for our analysis: $(1)$ how the DCC of the TMSV state behaves under varying noise strengths, and $(2)$ study a class of states, which contains the TMSV, in the ideal scenario. We show that, while more susceptible to the destructive effects of the considered decoherence noise models, the TMSV state can still maintain the highest capacity, while there does exist a class of pure Gaussian states which can furnish equivalent quantum advantage in the noiseless regime. This demonstrates that while we can opt for different states in the ideal situation, under imperfections in the protocol, the TMSV state is the way to go in order to successfully conduct dense coding with CV systems. 

\subsection{More quantum advantage with less robustness}
\label{subsec:TMSV_best}

\begin{figure*}[ht]
\includegraphics[width=\linewidth, height=10cm]{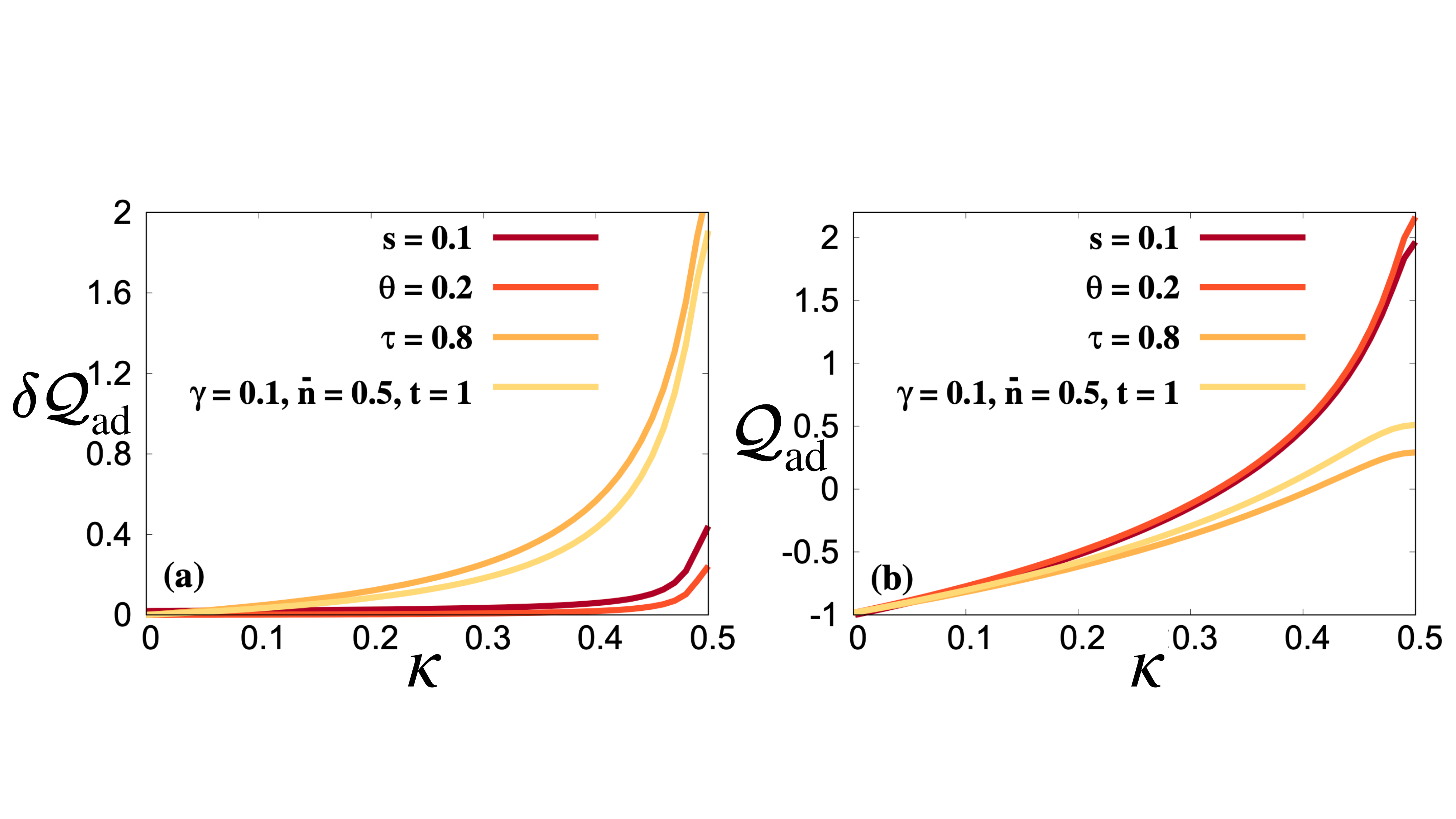}
\captionsetup{justification=Justified,singlelinecheck=false}
\caption{\textbf{Depletion in quantum advantage and the adaptive quantum advantage.} (a) The  change in the adaptive quantum advantage, $\delta \mathcal{Q}_{\text{ad}}$ (ordinate) as defined in the main text with the state parameter, $\kappa$ (abscissa) when affected by different forms of noise. From dark to light lines, we represent the quantum limited amplifier channel with $s = 0.1$ acting during state distribution, the pure-loss channel with $\theta = 0.2$ after encoding, environmental noise acting on the shared and encoded state with $\gamma = 0.1, \bar{n} = 0.5, t = 1$, and imperfect homodyne detection at $\tau = 0.8$ respectively. (b) The adaptive quantum advantage, $\mathcal{Q}_{\text{ad}}$ (ordinate) against $\kappa$ (abscissa) for the same specifications as in (a). The x-axis is dimensionless the y-axis is in bits.}
\label{fig:robust} 
\end{figure*}

Having focused on the TMSV state to investigate the effect of noise on the CV dense coding protocol, the question arises as to how the TMSV state fares if its generation itself suffers from imperfections.
Note that the TMSV state can be created when two modes, squeezed in the position and momentum quadratures respectively, are passed through a BS of transmissivity $\kappa = 0.5$. If the transmissivity varies, i.e,  $(0\leq \kappa \leq 0.5)$, we get a class of pure states, which we refer to as the $\kappa-$class, and whose bipartite entanglement is a monotonic function of $\kappa$, becoming maximum for the TMSV state.

Towards addressing the question of the robustness of a shared resource against noise, let us consider the TMSV state and the $\kappa-$ class of states and compare their DCC without and with noise. Quantitatively, we introduce the parameter, $\delta \mathcal{Q}=\mathcal{Q}_{\text{noiseless}}-\mathcal{Q}_{\text{noise}}$, which measures the robustness of a shared state against noise. Comparing two states, say $1$ and $2$, if we find $\delta \mathcal{Q}_1 > \delta \mathcal{Q}_2$, it implies that state $2$ is more robust in the presence of noise than state $1$. \ms{Note that, by robustness, we refer to the rate of degradation of the DCC: a higher value of $\delta \mathcal{Q}$ implies that $\mathcal{Q}_{\text{noise}}$ is much lower than in the noiseless case, i.e., the corresponding state is affected by imperfections to a much greater extent and is hence less robust.} The covariance matrix for the $\kappa-$class of states is given as
\begin{widetext}
\begin{equation}
    \Xi=\begin{pmatrix}
    e^{-2 r} (1 + (-1 + e^{4 r}) \kappa) & 0 & 2 \sqrt{-((-1 + \kappa) \kappa)} \sinh{2 r} &0\\
0 & e^{2 r} - 2 \kappa \sinh{2 r} & 0 & -2 \sqrt{-((-1 + \kappa) \kappa)} \sinh{2 r}\\
2 \sqrt{-((-1 + \kappa) \kappa)} \sinh{2 r} & 0 & e^{2 r} - 2 \kappa \sinh{2 r} &0\\
0 & -2 \sqrt{-((-1 + \kappa) \kappa)} \sinh{2 r} & 0 &  e^{-2 r} (1 + (-1 + e^{4 r}) \kappa)
    \end{pmatrix}.
\end{equation}
\end{widetext}
\textcolor{black}{ When we consider noise to be affecting the state during distribution, after encoding as well as through imperfect homodyne detection,} the mutual information of this class of states reads

\begin{widetext}
\begin{eqnarray}
    && \mathcal{I}^{\kappa}=\frac{1}{2} \log \Big(
     1 + \nonumber\\ 
     && \frac{2 \bar{N_s} - (-1 + x_3^2 y_1 + y_3) \tau - x_1^2 x_3^2 \tau \cosh{2 r}}{2 + \tau (-2 + y_2 + y_3 + x_2^2 (e^{2 r} - 2 \kappa \sinh{2 r}) + x_3 (e^{-2 r} (1 + (-1 + e^{4 r}) \kappa) x_1^2 x_3 + x_3 y_1 - 4 (\sqrt{(1 - \kappa) \kappa}) x_1 x_2 \sinh[2 r]))} \Big) + \nonumber \\
    && \frac{1}{2} \log \Big(1 + \nonumber \\ 
&& \frac{2 \bar{N_s} - (-1 + x_3^2 y_1 + y_3) \tau - x_1^2 x_3^2 \tau \cosh{2 r}}{2 + \tau (-2 + e^{-2 r} (1 + (-1 +    e^{4 r}) \kappa) x_2^2 + y_2 + y_3 + x_3 (x_3 y_1 - 4 (\sqrt{(1 - \kappa) \kappa}) x_1 x_2 \sinh{2 r} + x_1^2 x_3  (e^{2 r} - 2 \kappa \sinh{2 r})))} \Big).\nonumber\\
   \label{eq:mu-new}
\end{eqnarray}
\end{widetext}
In the adaptive scheme, we are unable to determine $r^{\text{opt}} $ analytically from Eq. \eqref{eq:mu-new}  by maximizing over $r$, and hence, we perform numerical optimization over $r$ for fixed noise parameters and fixed energy $(\bar{N_s}=30)$, resulting in the DCC. For presenting the observation, we opt for the limited amplifier noise at $s=0.1$, a pure loss channel at $\theta=0.2$, and an environmental noise at $\gamma=0.1$,~$\bar{n}=0.5$ and $t=1$, affecting both shared and encoded states. For the imperfect double-homodyne, we will take $\tau=0.8$.

We observe that although $\delta \mathcal{Q}^{\text{TMSV}} >\delta \mathcal{Q}^{\kappa} $ for all other values of $\kappa \neq 0.5$, implying less robustness for the TMSV state as compared to other states in this class, $\mathcal{Q}^{TMSV}_{\text{ad}} >\mathcal{Q}^{\kappa}_{\text{ad}} $, i.e., the quantum advantage for the TMSV state even in presence of noise remains higher than the states with $\kappa \neq 0.5$ (as shown in Fig. \ref{fig:robust}). This qualitative discussion is valid even for other choices of parameters in the noise models and \ms{demonstrates the peculiar superiority offered by the TMSV state over other Gaussian states in CV dense coding routines.}

\subsection{A class of pure Gaussian states with maximum dense coding capacity}
\label{subsec:pure_dcc}

In this penultimate section, we shall demonstrate that all pure states that have a specific form can provide the same dense coding capacity as the TMSV state with equal energy. Let us consider a class of pure \ms{Gaussian} states that are represented by the following covariance matrix

\begin{equation}
\Xi_{\text{pure}}=\begin{pmatrix}
aI_2 & \sqrt{a^2 - 1}\sigma_z\\
\sqrt{a^2 - 1}\sigma_z & aI_2
\end{pmatrix}
\label{eq:initial_state_pure} ,
\end{equation}
where the displacement vector may be taken to be $\mathbf{d}_{\text{pure}} = (0, 0, 0, 0)^T$ without loss of generality. Note that the TMSV state corresponds to $a = \cosh 2r$ in this representation.

Upon encoding at the sender's mode with a Gaussian probability distribution of standard deviation, $\sigma$, the mutual information can be written as

\begin{equation}
    \mathcal{I}^{\text{pure}} = \log \Big(1 + 2 \sigma^2 (a + \sqrt{a^2 - 1}) \Big),
    \label{eq:MI_pure}
\end{equation}
while the sender's side energy has the form $\bar{N}_{S-\text{pure}} = \frac{1}{2}(a + 4\sigma^2 - 1)$. Substituting $\sigma$ in Eq. \eqref{eq:MI_pure} and optimizing with respect to the state parameter $a$ leads to $a^{\text{opt}} = \frac{(1 + 2 \bar{N}_{S-\text{pure}})^2 + 1}{2 (1 + 2 \bar{N}_{S-\text{pure}})}$ and the corresponding DCC can be found to be $\mathbf{C}_{\text{pure}} = \log (1 + \bar{N}_{S-\text{pure}} + \bar{N}_{S-\text{pure}}^2)$ which is exactly the same as that of a TMSV state having average energy at the sender's mode equal to $\bar{N}_{S-\text{pure}}$. \textcolor{black}{Therefore, all states, characterized by Eq. \eqref{eq:initial_state_pure}, lead to the \ms{same maximal dense coding capacity among Gaussian states for a given energy at the sender's mode.}}

This observation may be explained by the entanglement content of the states.  In the case of the TMSV state, the sender's side energy, $\bar{N}_{S-\text{TMSV}} = \sinh^2 r + 2 \sigma^2$, and $\nu_{\text{TMSV}} = \cosh 2r = 2\bar{N}_{S-\text{TMSV}} + 1 - 4 \sigma^2$ required to compute entanglement \cite{Serafini_2017} while  $\bar{N}_{S-pure} = \frac{a - 1 + 4 \sigma^2}{2}$ and the symplectic eigenvalue $\nu_{\text{pure}} = a = 2 \bar{N}_S + 1 - 4 \sigma^2$, which is exactly the same relation as that of the TMSV state. Therefore, pure \ms{Gaussian} states of the considered form have the same degree of entanglement as a TMSV state of equivalent energy. This indicates why such states can also provide the maximal dense coding capacity \ms{among Gaussian states employed in the protocol}. This phenomenon is further vindication of our observation (see Appendix \ref{app:numerics}) that the Holevo quantity of pure states is a monotonically increasing function of entanglement at a fixed energy of the sender's mode.


Furthermore, our numerical simulations reveal that when states of the form given by Eq. \eqref{eq:initial_state_pure} are affected by noise, the resulting quantum advantage is exactly the same as that of the equal-energy noisy TMSV state. This reveals that such states have the same robustness against noise as the TMSV state. \ms{Therefore, the TMSV state is not the only maximal Gaussian resource} in CV dense coding, but there exists a wider class of \ms{Gaussian} states that can furnish the maximum dense coding capacity. This reflects the fact that, similar to discrete variable DC, all maximally entangled pure \ms{Gaussian} states at fixed energy provide \ms{maximal DCC (among Gaussian states)}, even though the continuous variable DCC is not in one-to-one correspondence with entanglement for pure states.

\section{Conclusion}
\label{sec:conclu}

The dense coding protocol, used for the transmission of classical information between distant parties, has been widely realized in optical setups, and continuous variable (CV) states are proven to be an efficient resource towards obtaining non-classical dense coding capacities. The original CV dense coding protocol was proposed with the two-mode squeezed vacuum (TMSV) state. However, during the physical implementation of the protocol, noise is inevitable and is expected to adversely impact the success of the scheme. Noise can affect the resource state in several instances, e.g., during the distribution of the state between the involved parties and also during the transmission of the encoded mode from the sender to the receiver. There can also exist imperfections in the decoding setup, which would further lower the regime of quantum advantage.

In this work, we analyzed the dense coding protocol between a single sender and a single receiver in the presence of noise and imperfections. In particular, we found the expression for the mutual information of a generic two-mode Gaussian state when it is affected by noise represented by deterministic Gaussian completely positive maps. To compute the dense coding capacity, we considered two independent schemes - \ms{the adaptive scheme, and the non-adaptive scheme.}

For illustration, we considered two kinds of noisy channels - the quantum-limited amplifier and the pure-loss channel - through which the modes are transmitted, and calculated the dense coding capacity for each noisy channel. The channels were assumed to act on the resource state both before and after the encoding process.
Moreover, we considered an optical design to model the imperfect double-homodyne detection to study the impact of erroneous decoding on the dense coding capacity. We demonstrated that the adaptive scheme always furnishes a higher dense coding capacity and allows for quantum advantage over a larger range of noise strength. Interestingly, we showed that the nature of the pure-loss channel allows the dense coding capacity to revive at periodic instances, thereby making the quantum advantage attain its noiseless limit. Considering the TMSV state as the resource, we provided threshold noise strengths beyond which non-classical capacity can be obtained for both kinds of noises. Using our proposed model for faulty double-homodyne detection, we also quantified the range of imperfection which allows for quantum advantage even when the decoding mechanism is not ideal. Apart from imperfect quantum channels, we also studied the effect of environmental interaction on the dense coding capacity furnished by the resource state. We demonstrated that a greater coupling strength between the system and the environment, and a higher mean photon number of the environment, led to a faster collapse of the quantum advantage. Furthermore, the impact of the coupling strength was shown to be much more detrimental than that of the mean photon number of the environment. A curious result we reported is that the TMSV state provides the best quantum advantage under noise and imperfections, even though its capacity is the most depleted in the presence of noise. This demonstrates the superiority of the TMSV state in the dense coding protocol despite its higher fragility than the other states having lower entanglement.

The conditional entropy of the resource state has been established as a resource for quantum advantage in dense coding in the discrete variable case, although this result is not apparent for continuous variable systems. 
Here, \textcolor{black}{we numerically demonstrated that it can indeed serve as a precursor for quantum advantage in Gaussian CV dense coding protocols.} Specifically, a positive difference between the conditional entropy of the state and its value at the threshold energy mimics the quantum advantage. Furthermore, the higher the conditional entropy difference, the higher the quantum advantage over the classical information transmission scheme. 
We also identified another class of pure \ms{Gaussian} states that has identical dense coding capacity as the TMSV state with equal energy both in the noiseless and noisy scenarios. We note that any pure state can be brought to the considered form through local symplectic operations that leave the entanglement invariant, thereby providing a wider selection of resource states for obtaining quantum advantage in CV dense coding routines.


\section*{Acknowledgement} 
We acknowledge the support from the Interdisciplinary Cyber-Physical Systems (ICPS) program of the Department of Science and Technology (DST), India, Grant No.: DST/ICPS/QuST/Theme- $1/2019/23$ 
and  ``Technology Vertical - Quantum Communication'' under the National Quantum Mission of the Department of Science and Technology (DST)  (Sanction Order No. DST/QTC/NQM/QComm/$2024/2$ (G)).
We acknowledge the use of QIClib -- a modern C++ library for general-purpose quantum information processing and quantum computing~\cite{QICLib}.  This research was supported in part by the 'INFOSYS scholarship for senior students'. R.G. acknowledges funding from the HORIZON-EIC-$2022$-PATHFINDERCHALLENGES-$01$ program under Grant Agreement No.~$10111489$ (Veriqub).  Views and opinions expressed are those of the authors only and do not necessarily reflect those of the European Union. Neither the European Union nor the granting authority can be held responsible for them.

\appendix

\section{The Holevo quantity for pure two-mode CV states}
\label{app:numerics}

\begin{figure}[ht]
\includegraphics[width=\linewidth, height=5.8cm]{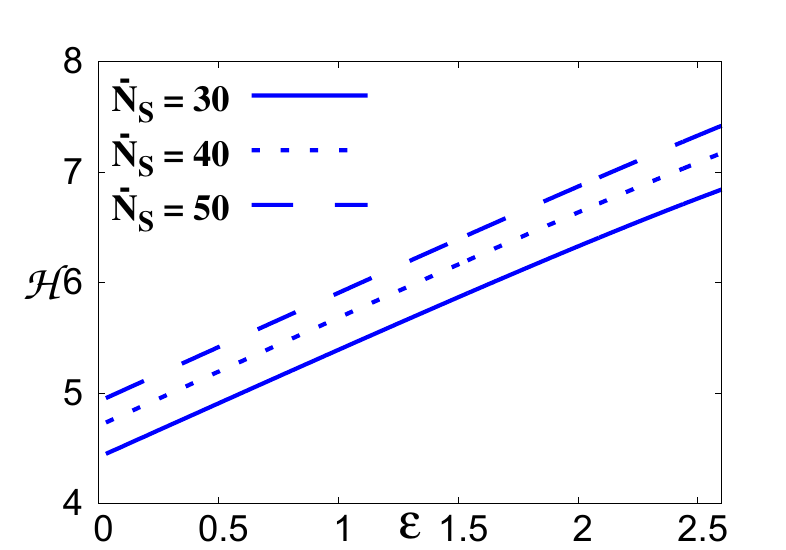}
\captionsetup{justification=Justified,singlelinecheck=false}
\caption{\textbf{The Holevo quantity for pure two-mode Gaussian states.} The variation of the Holevo quantity, $\mathcal{H}$ (ordinate) against entanglement, $\mathcal{E}$ (abscissa), for pure two-mode Gaussian states with sender-side energy, $\bar{N}_S = 30$ (solid line), $\bar{N}_S = 40$ (dotted line), and $\bar{N}_S = 50$ (dashed line). The horizontal axis is in ebits while the vertical axis is in bits.}
\label{fig:holevo} 
\end{figure}

Let us recall that in the case of discrete variable DC, the capacity of pure states is uniquely determined by their entanglement. Such a relation, however, is not straightforward to assume for dense coding with continuous variable states. While it is not analytically possible to estimate the dense coding capacity of arbitrary pure Gaussian states, we can still determine the Holevo quantity given by 

\begin{equation}
    \mathcal{H} = S(\int d^2 \alpha P(\alpha) \rho_\alpha) - \int d^2 \alpha S(\rho_\alpha),
    \label{eq:holevo}
\end{equation}
where $\rho_\alpha$ is the encoded state, with the encoding probability distribution over the displacement parameter being given by $P(\alpha)$. Estimating Eq. \eqref{eq:holevo} for pure two-mode Gaussian states, we can arrive at the following observation  \textbf{:}

\textbf{Observation.} \textit{The Holevo quantity of all pure two-mode Gaussian states having the same sender-side energy is proportional to the entanglement.}

To verify our claim, we simulate $10^4$ random two-mode pure Gaussian states~\cite{Roy_arXiv_2024} with different sender-side energies. From Fig. \ref{fig:holevo}, the one-to-one correspondence between the Holevo quantity and the entanglement of the states having a fixed sender-side energy is apparent.

Let us now elucidate the procedure adopted for simulating random two-mode pure Gaussian states, $\ket{\psi}_{AB}$, for which we try to estimate the Holevo quantity and the corresponding entanglement. We note that, since the states are pure, the Holevo quantity in Eq. \eqref{eq:holevo} reduces to $\mathcal{H} = S(\int d^2 \alpha P(\alpha) \rho_\alpha)$ while the entanglement is quantified by the von-Neumann entropy of the reduced single-mode state, i.e., $S(\rho_{A})= -\Tr(\rho_A \log \rho_A)$ where $\rho_A = \Tr_B \ketbra{\psi}{\psi}_{AB}$ is the reduced subsystem of at the sender's side and $S(x)=-\Tr(x \log x)$ is the von Neumann entropy. Specifically, for a Gaussian state, the entanglement reduces to $\epsilon = \frac{\nu+1}{2}\log_2\frac{\nu+1}{2} - \frac{\nu-1}{2}\log_2\frac{\nu-1}{2}$, where $\nu$ represents the symplectic eigenvalue of the covariance matrix of the reduced state. Since the von Neumann entropy depends only on the symplectic eigenvalues of the covariance matrix of the Gaussian state, we shall only be concerned with generating random two-mode covariance matrices while assuming that the displacement vector is given by $\mathbf{d} = (0, 0, 0, 0)^T$.

The covariance matrix of any pure two-mode Gaussian state may be represented as~\cite{Fukuda_JMP_2019, Roy_arXiv_2024}

\begin{equation}
    \Xi = O \Gamma O^T,
    \label{eq:random_cov-mat}
\end{equation}
where $O$ represents an orthogonal symplectic matrix and $\Gamma = \mathcal{D} \oplus \mathcal{D}^{-1}$. The energy of the state is completely specified by $\Gamma$, and for a given sender-side energy of $\bar{N}_S$, we choose $\mathcal{D} = \frac{\bar{N}_S}{2} I_2$~\cite{Roy_arXiv_2024}. Then, for such fixed energy, the random covariance matrices are simulated by randomly generating $O$ as~\cite{Roy_arXiv_2024}

\begin{equation}
    O = \begin{pmatrix}
        Re[U] & Im [U]\\
        -Im[U] & Re[U]
    \end{pmatrix},
    \label{eq:ortho_symp}   
\end{equation}
where $U \in SU(2)$ is a random unitary matrix while $Re$ and $Im$ denote real and imaginary parts respectively. In our calculations, we fix $\bar{N}_S = 30, 40, 50$ and for each case, we choose $10^4$ random orthogonal symplectic matrices according to Eq. \eqref{eq:ortho_symp} in order to generate random two-mode pure Gaussian states.

The encoding probability distribution is taken to be $P(\alpha) = \frac{1}{2 \pi \sigma^2} \exp [-(\alpha_x^2 + \alpha_p^2)/2\sigma^2] = \frac{1}{4 \pi \sigma^2} \exp [-(x^2 + p^2)/4\sigma^2]$, with $\alpha = \alpha_x + \iota \alpha_y = (\langle \hat{x} \rangle +  \iota \langle \hat{p} \rangle)/\sqrt{2}$. The covariance matrix corresponding to the encoded ensemble in the Holevo quantity may be estimated as~\cite{Serafini_2017}

\begin{equation}
    \int d^2 \alpha P(\alpha) \rho_\alpha \equiv \Xi + 4 \sigma^2 \begin{pmatrix}
        I_2 & 0 \\
        0 & 0
    \end{pmatrix}.
    \label{eq:classical_mix}
\end{equation}
Now, finding the symplectic eigenvalues, $\nu_i$, of Eq. \eqref{eq:classical_mix} allows us to estimate the Holevo quantity whereas those corresponding to the covariance matrix of the sender-mode give us the entanglement. In our calculations, we set $\sigma = 1$ which makes the sender-side energy $\bar{N}_S + 2$.

\textcolor{black}{\section{Highest dense coding capacity for the TMSV state} \label{app:tmsv_cap}}

Throughout our work, we have considered two-mode Gaussian states which are characterized by a covariance matrix of the form
\begin{equation}
    \Xi=\begin{pmatrix}
\mathcal{A} & \mathcal{B}\\
\mathcal{B} & \mathcal{C}
\end{pmatrix},
\end{equation}
as defined in Eq. \eqref{eq:initial_state}. In this section, we aim to prove that among all such \ms{Gaussian} states, the TMSV state furnishes the highest dense coding capacity. To that end, we note that such states can be prepared by two single-mode squeezing operations followed by a global two-mode squeezing on two vacuum states~\cite{Tserkis_PRA_2017}, i.e.,

\begin{equation}
    \Xi = S_2(r) \Big(S_1(s_1) \oplus S_1(s_2) \Big) I_4 \Big(S_1(s_1) \oplus S_1(s_2) \Big)^T S_2(r)^T,
    \label{eq:two-mode_decomposition}
\end{equation}
where $S_1(s_i)$ and $S_2(r)$ are the symplectic transformations corresponding to single- and two-mode squeezing parameters, respectively, with squeezing degrees $s_i$ and $r$ and $I_4$ is the covariance matrix of the two-mode vacuum state. Note that the single-mode squeezing parameters, $s_i$ comprise a squeezing degree, $\tilde{s}_i$, together with a squeezing angle, $\theta_i$, i.e., $s_i = \tilde{s}_i e^{\iota \theta_i}$. The TMSV state is obtained with $\tilde{s}_i = 0$ for $i = 1, 2$.

Let us first concentrate on the entanglement content of the considered pure state, at a given sender-side energy. The symplectic eigenvalue of the reduced state is given by $\nu = \cosh 2r$ whereas the energy of the sender mode reads as $\bar{N}_S = (\cosh 2r \cosh 2 \tilde{s}_1 - 1)/2$. To obtain the entanglement as a function of the energy, we recast the symplectic eigenvalue as $\nu = (2 \bar{N}_S + 1)/\cosh 2\tilde{s}_1$. It is immediately evident that $\nu$ decreases with an increase in $\tilde{s}_1$ and since the entanglement is a monotonic function of the symplectic eigenvalue, the maximum entanglement at a given energy is obtained at $\tilde{s}_1 = 0$. Therefore, in order to calculate the capacity, we shall set $\tilde{s}_1 = 0$, since we have already proved that the Holevo quantity of pure states is proportional to their entanglement at a fixed $\bar{N}_S$. Furthermore, the entanglement, being independent of the squeezing angle, we can safely choose $\theta_1 = 0$ and $\theta_2 = \pi$ without loss of generality.

The mutual information for the state in Eq. \eqref{eq:two-mode_decomposition} with $\tilde{s}_1 = 0, \theta_1 =0, \theta_2 = \pi$, in terms of the sender-side energy, is given as 

    \begin{eqnarray}
        \mathcal{I} = \frac{1}{2} \log (1 + e^{2 r} \zeta ) \Big(1 + e^{2 (r + \tilde{s}_2)} \zeta \Big),
        \label{eq:app_mi}
    \end{eqnarray}
where, $\zeta = (\cosh^2 r + \sinh^2 r \cosh 2 \tilde{s}_2 - 2 \bar{N}_S - 1)(\tanh \tilde{s}_2 - 1)$. Taking the derivative of Eq. \eqref{eq:app_mi} with respect to $r$ and $\tilde{s}_2$, we simultaneously solve the following two equations \textbf{:}

\begin{eqnarray}
  &&  \cosh 2 \tilde{s}_2 (e^{2r} - 1) + e^{2r} - 4 \bar{N}_S - 1 = 0, ~ \text{and} ~ \\
  && \nonumber  \sinh^2 r  \sinh 2 \tilde{s}_2 (4 \sinh 2 r - 3 \cosh 2 r - 1) \\
  && - 4 \bar{N}_s^2 \tanh \tilde{s}_2 \sech^2 \tilde{s}_2 = 0.
\end{eqnarray}
The solutions for optimal $r$ and $\tilde{s}_2$ yield $r_{\text{opt}} = \frac{1}{2} \log (1 + 2 \bar{N}_S)$ and $\tilde{s}_{2-\text{opt}} = 0$, which correspond to the TMSV state (since $\tilde{s}_1 = 0$ already). Therefore, in the noiseless regime, the TMSV state provides the highest dense coding capacity for \ms{Gaussian} states of the form considered in this manuscript.\\

\section{Non-adaptive dense coding protocol with generic Gaussian states under different noise models}
\label{app:nonadaptive_dc}

\textcolor{black}{Here we present numerical evidence suggesting that the negative conditional von Neumann entropy \ms{remains an indicator of quantum advantage} in the non-adaptive noisy dense coding protocol involving Gaussian states. In this case, the encoding strategy is fixed and independent of the channel acting on the state. Quantum advantage is again observed only beyond a threshold value of the negative conditional entropy. Specifically, for the pure loss channel, the non-adaptive quantum advantage becomes positive beyond $-\tilde S(A|B)_{\text{th}} \simeq 1.6403$ and $1.1045$ at $\theta=0.1$ and $0.4$, respectively, and increases monotonically with the sender’s energy ${\bar N}_S$ (see Fig.~\ref{fig:non_adaptive_general_all}(a)). A similar behavior is found for the quantum-limited amplifier channel, with threshold values $-\tilde S(A|B)_{\text{th}}\simeq 1.619$ and $1.187$ for $s=0.1$ and $0.4$ (see Fig.~\ref{fig:non_adaptive_general_all}(b)). These results confirm that, even without adaptive optimization, negative conditional entropy remains a sufficient indicator of quantum advantage in Gaussian CV dense coding.} 

\begin{figure*}[htb!]
\includegraphics[width=\linewidth]{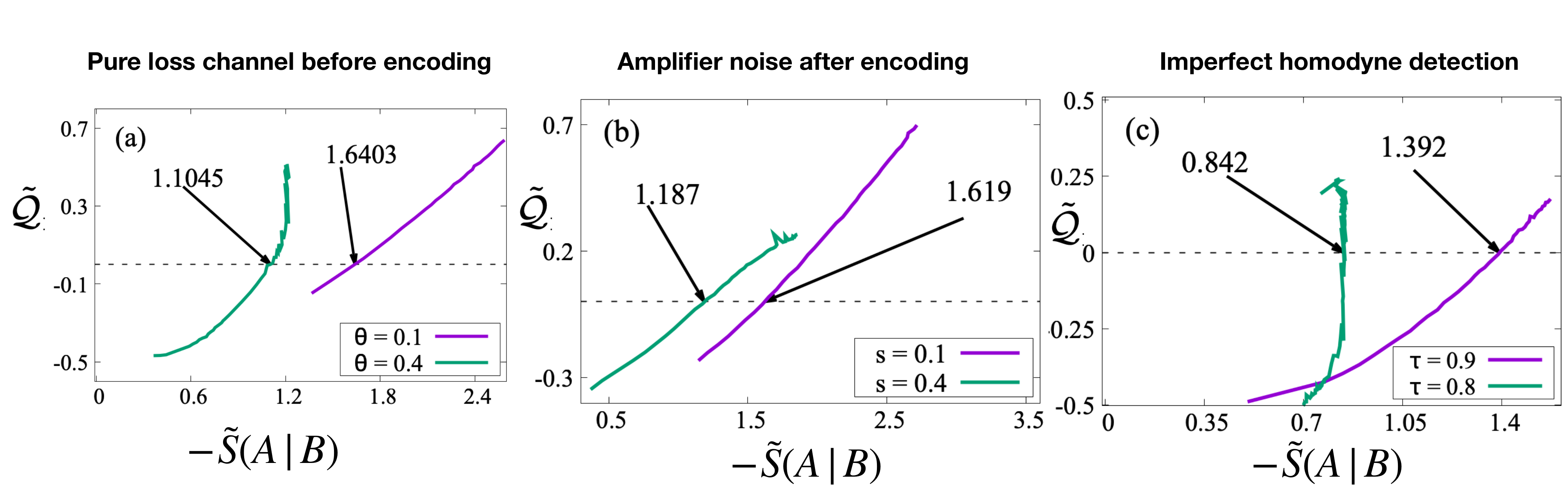}
\captionsetup{justification=Justified,singlelinecheck=false}
\caption{\textcolor{black}{\textbf{Quantum advantage and negative conditional entropy in the non-adaptive scenario for the state of Eq.~(\ref{eq:initial_state}), with imperfections.}  The variation of the quantum advantage, $\tilde{\mathcal{Q}}=(\tilde{\mathbf{C}}_{\text{non-ad}} - \mathbf{C}_{\text{cl}} )$ (ordinate), as a function of the negative conditional entropy, $-\tilde{S}(A|B)$ (abscissa), for the general state in Eq.~(\ref{eq:initial_state}). Panels (a) and (b) show the cases of a pure-loss channel before encoding and a quantum-limited amplifier after encoding, with noise strengths $s(\theta)=0.1$ and $0.4$, shown in purple (dark) and green (light), respectively. Panel (c) depicts the effect of imperfect homodyne detection for $\tau=0.9$ and $0.8$, again in purple (dark) and green (light), respectively. Both axes are expressed in bits.
 }}
\label{fig:non_adaptive_general_all} 
\end{figure*}

\bibliographystyle{apsrev4-2}
	\bibliography{ref}

\end{document}